\documentclass[aps,prd,twocolumn,10pt,superscriptaddress,nofootinbib,nobibnotes,longbibliography]{revtex4-1}

\usepackage{amssymb}
\usepackage{graphicx}
\usepackage{amsmath}
\usepackage{hyperref}
\usepackage{subfigure}
\usepackage{multirow}
\usepackage{setspace}
\usepackage{verbatim}
\usepackage{float}
\usepackage{color}
\usepackage{ulem}
\usepackage[utf8]{inputenc}
\usepackage[table,xcdraw]{xcolor}
\usepackage{makecell}

\begin{document}

\title{Primordial black holes and curvature perturbations from false vacuum islands}

\author{Rong-Gen Cai}
\email{cairg@itp.ac.cn}
\affiliation{School of Physical Science and Technology, Ningbo University, Ningbo, 315211, China}
\affiliation{CAS Key Laboratory of Theoretical Physics, Institute of Theoretical Physics, Chinese Academy of Sciences (CAS), Beijing 100190, China}

\author{Yu-Shi Hao}
\email{haoyushi@itp.ac.cn}
\affiliation{CAS Key Laboratory of Theoretical Physics, Institute of Theoretical Physics, Chinese Academy of Sciences (CAS), Beijing 100190, China}
\affiliation{School of Physical Sciences, University of Chinese Academy of Sciences (UCAS), Beijing 100049, China}

\author{Shao-Jiang Wang}
\email{schwang@itp.ac.cn (corresponding author)}
\affiliation{CAS Key Laboratory of Theoretical Physics, Institute of Theoretical Physics, Chinese Academy of Sciences (CAS), Beijing 100190, China}
\affiliation{Asia Pacific Center for Theoretical Physics (APCTP), Pohang 37673, Korea}


\begin{abstract}
Recently, much attention has been focused on the false vacuum islands that are flooded by an expanding ocean of true-vacuum bubbles slightly later than most of the other parts of the world. These delayed decay regions will accumulate locally larger vacuum energy density by staying in the false vacuum longer than those already transited into the true vacuum.  A false vacuum island with thus acquired density contrast of a super-horizon size will evolve locally from radiation dominance to vacuum dominance, creating a local baby Universe that can be regarded effectively as a local closed Universe. If such density contrasts of super-horizon sizes can ever grow large enough to exceed the threshold of gravitational collapse, primordial black holes will form similar to those collapsing curvature perturbations on super-horizon scales induced by small-scale enhancements during inflation. If not, such density contrasts can still induce curvature perturbations potentially observable today. In this paper, we revisit and elaborate on the generations of primordial black holes and curvature perturbations from delayed-decayed false vacuum islands during asynchronous first-order phase transitions with fitting formulas convenient for future model-independent studies.
\end{abstract}
\maketitle
 
\section{Introduction}\label{sec:introduction}

The cosmological first-order phase transition (FOPT) \cite{Mazumdar:2018dfl,Hindmarsh:2020hop,Caldwell:2022qsj,Athron:2023xlk} plays an essential role in probing the new physics~\cite{Cai:2017cbj,Bian:2021ini} beyond the standard model of particle physics in the early Universe via future detection of stochastic gravitational wave (GW) backgrounds~\cite{Caprini:2015zlo,Caprini:2019egz,LISACosmologyWorkingGroup:2022jok} in the space-borne GW detectors like LISA~\cite{Armano:2016bkm,Audley:2017drz,LISA:2022yao} and Taiji~\cite{Hu:2017mde,Ruan:2018tsw,TaijiScientific:2021qgx} as well as TianQin~\cite{TianQin:2015yph,Luo:2020bls,TianQin:2020hid}. Recently, there has been a renewed interest in generating primordial black holes (PBHs) from cosmological FOPTs via shrinking~\cite{Baker:2021nyl,Kawana:2021tde} or accumulating~\cite{Liu:2021svg} mechanisms other than early proposals of colliding mechanism~\cite{Hawking:1982ga,Crawford:1982yz,Moss:1994pi,Moss:1994iq} (see recent studies~\cite{Freivogel:2007fx,Johnson:2011wt,Kusenko:2020pcg,Jung:2021mku}).

The shrinking mechanism~\cite{Baker:2021nyl,Kawana:2021tde} originated from the pictures of filtered dark matter~\cite{Baker:2019ndr} and Fermi-ball dark matter~\cite{Hong:2020est} during some specific  FOPTs, where some particle species can acquire too large masses to barely penetrate the true-vacuum bubbles but are reflected and then trapped in the false vacuum phase, leading to direct formations of PBHs~\cite{Baker:2021nyl} or indirect PBH formations via collapsing Fermi balls~\cite{Kawana:2021tde}. Note that the reflected would-be massive particles in the filtered dark matter scenario can also annihilate to trigger baryogenesis~\cite{Arakawa:2021wgz,Baker:2021zsf} instead of producing PBHs, and the Fermi-ball dark matter scenario can similarly produce baryon asymmetry via leptogenesis~\cite{Huang:2022vkf}. The direct shrinking mechanism~\cite{Baker:2021nyl} was further developed in Ref.~\cite{Baker:2021sno} and exemplified for a PeV-scale model~\cite{Cline:2022xhx}. The indirect shrinking mechanism~\cite{Kawana:2021tde} was further elaborated on the population statistics~\cite{Lu:2022paj} and its evolutionary endpoint~\cite{Kawana:2022lba} for the old phase remnants, and also exemplified for an electroweak model~\cite{Huang:2022him}. Correlated signals~\cite{Marfatia:2021hcp,Tseng:2022jta,Gehrman:2023esa,Xie:2023cwi,Banerjee:2023brn,Chen:2023oew,Borah:2024lml} were also proposed to discriminate different formation channels. Despite being highly model-dependent for both shrinking mechanisms, the direct shrinking mechanism was further criticized recently in Ref.~\cite{Lewicki:2023mik} for the backreaction from trapped particles to reduce the compactness needed for PBH formations.

On the other hand, the accumulating mechanism~\cite{Liu:2021svg} is model-independent, imposing no constraint on the properties of particles of phase transition models, but only takes advantage of inhomogeneous evolutions of asynchronous transition regions due to the stochastic nature of bubble nucleations. These delayed-decayed regions resemble and revive earlier proposals of trapped false vacuum domains~\cite{Sato:1981bf,Maeda:1981gw,Sato:1981gv,Kodama:1981gu,Sato:1981hk,Kodama:1982sf}~\footnote{See also a recent study~\cite{Caravano:2024tlp} simulating the non-perturbative dynamics of single field inflation that might offer alternative channel for PBH formation from trapped false vacuum regions during inflation.}, and their subsequent PBH formations were further exemplified in the QCD-scale~\cite{He:2022amv,He:2023ado,Gouttenoire:2023bqy,Salvio:2023blb}, electroweak-scale~\cite{Hashino:2021qoq,Hashino:2022tcs,Salvio:2023ynn,Gouttenoire:2023pxh,Conaci:2024tlc} and PeV-scale~\cite{Baldes:2023rqv} models. More analytic estimations on the parameter dependence were carried out recently in Refs.~\cite{Kawana:2022olo,Gouttenoire:2023naa,Lewicki:2023ioy} when these delayed-decayed false vacuum islands are approximated locally as exponentially expanding regions for an exponential nucleation rate~\cite{Kawana:2022olo,Gouttenoire:2023naa} or with a quadratic correction~\cite{Lewicki:2023ioy,Kanemura:2024pae}. In particular, different from earlier proposal~\cite{Kodama:1982sf} and its recent revisit~\cite{Lewicki:2023ioy} where only those pure false vacuum regions that never underwent single nucleation are supposed to collapse into PBHs~\cite{Kawana:2022olo}, our accumulating mechanism~\cite{Liu:2021svg} and its recent refinement~\cite{Gouttenoire:2023naa} allows causal regions to collapse into PBHs even after they already started the process of nucleation as long as they started slightly late enough to accumulate sufficient local overdensities. Moreover, the recent refinement~\cite{Gouttenoire:2023naa} further excludes bubble nucleations outside the postponed Hubble patches but inside the past light cone of final collapsed regions. Furthermore, another recent refinement~\cite{Lewicki:2023ioy} includes the bubble wall energy propagating into the postponed Hubble patch from those bubbles nucleated inside the past light cone of collapsed regions.

\begin{figure*}
    \centering
    \includegraphics[width=0.24\textwidth]{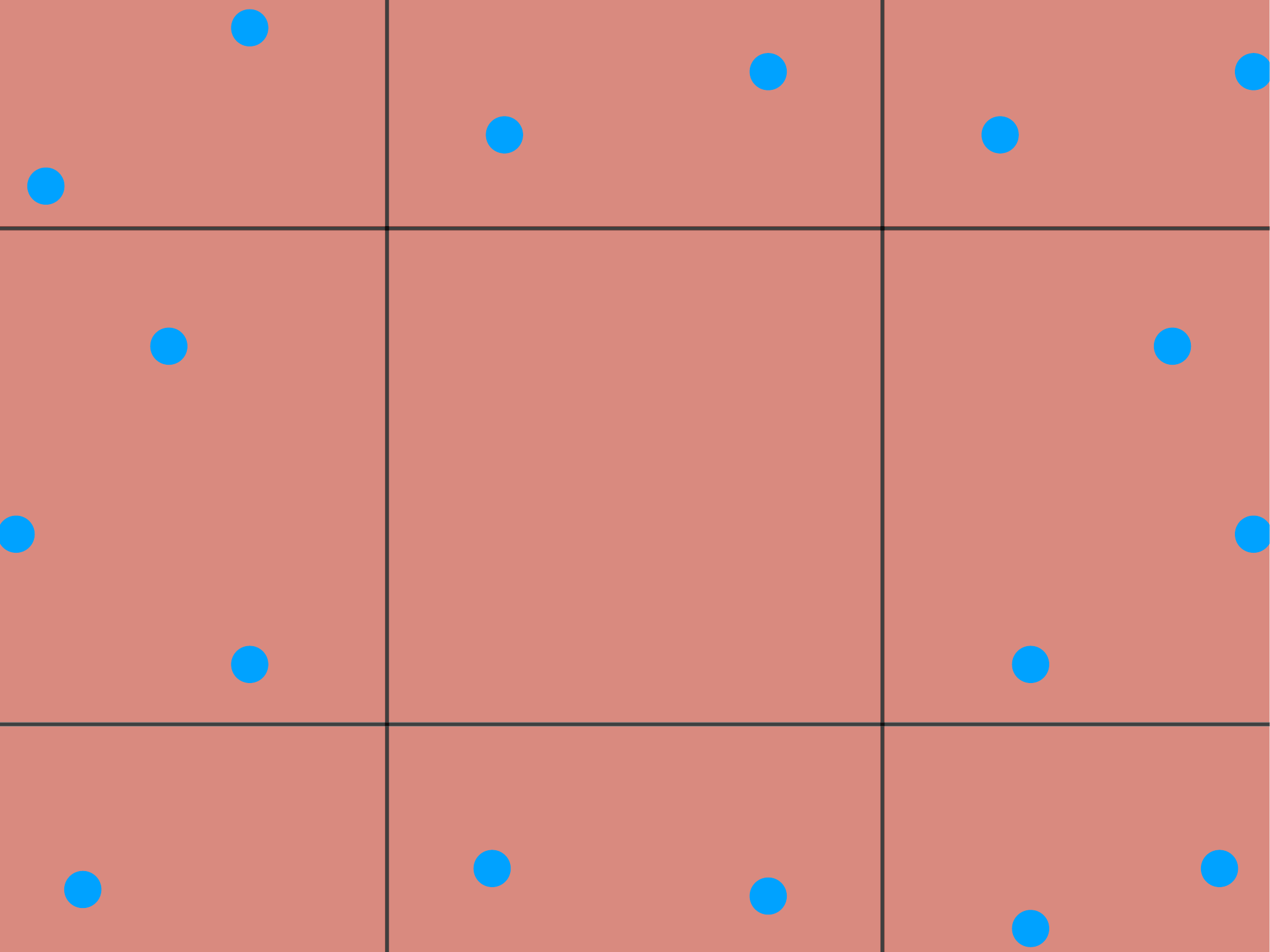}
    \includegraphics[width=0.24\textwidth]{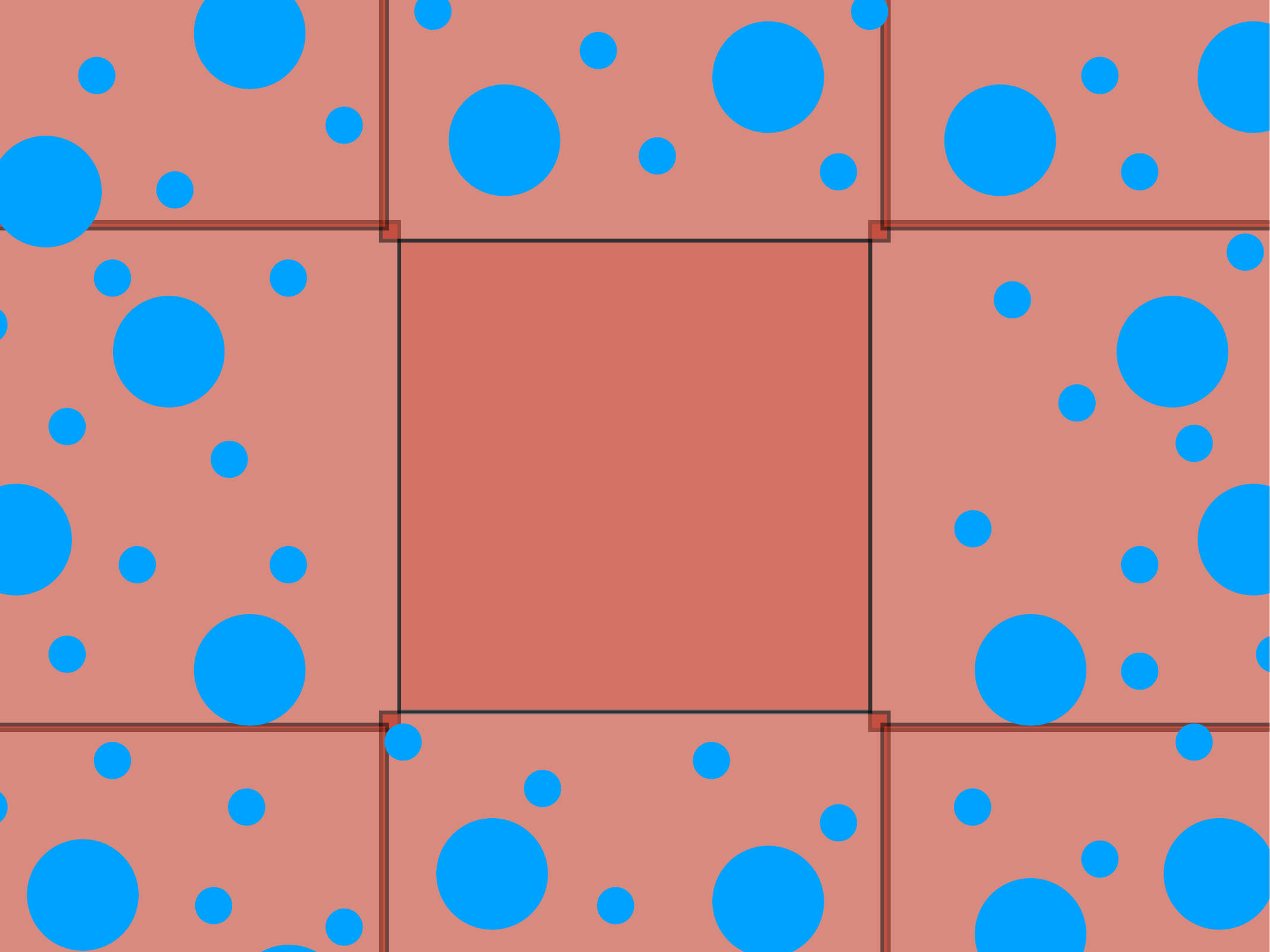}
    \includegraphics[width=0.24\textwidth]{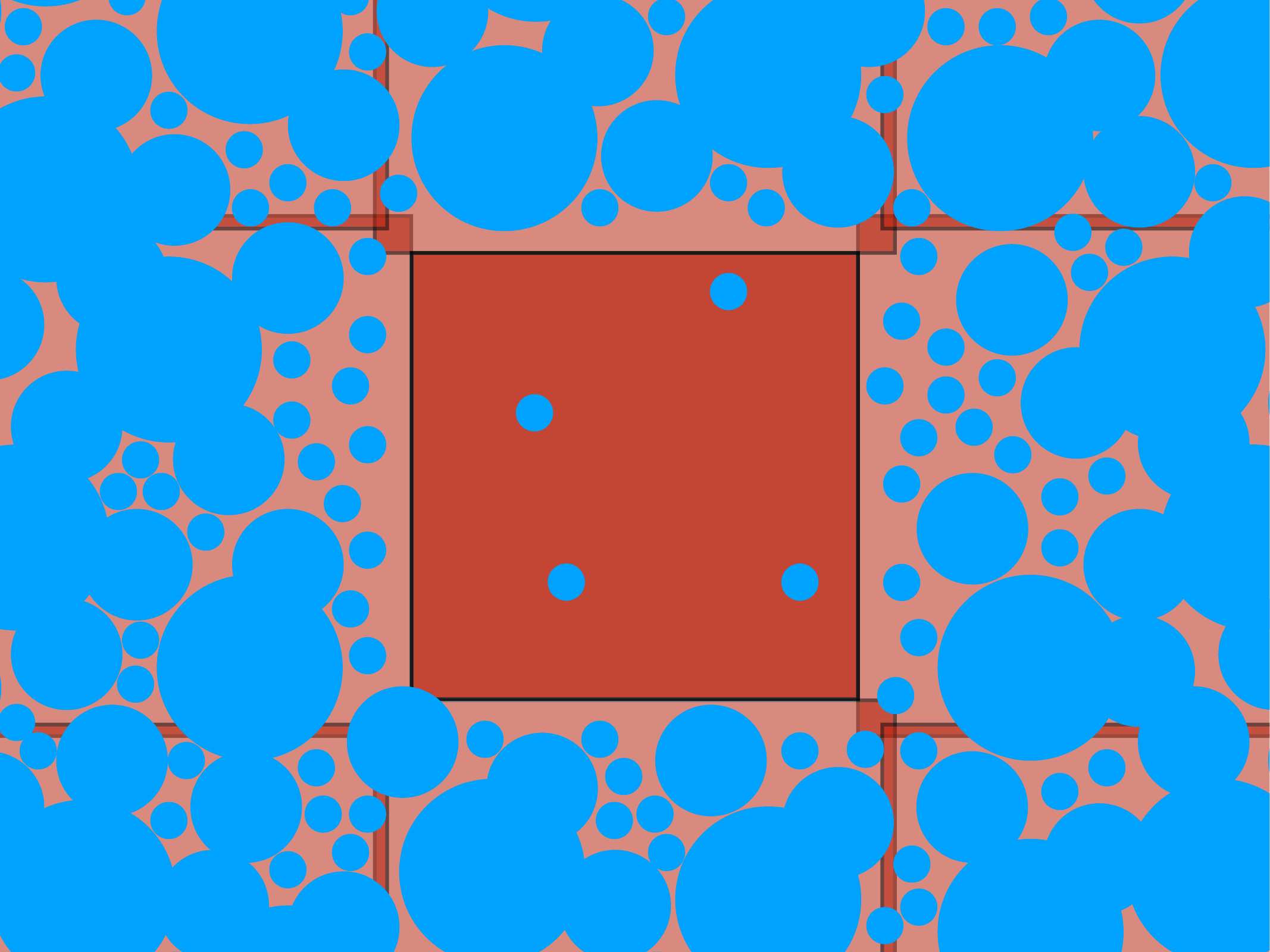}
    \includegraphics[width=0.24\textwidth]{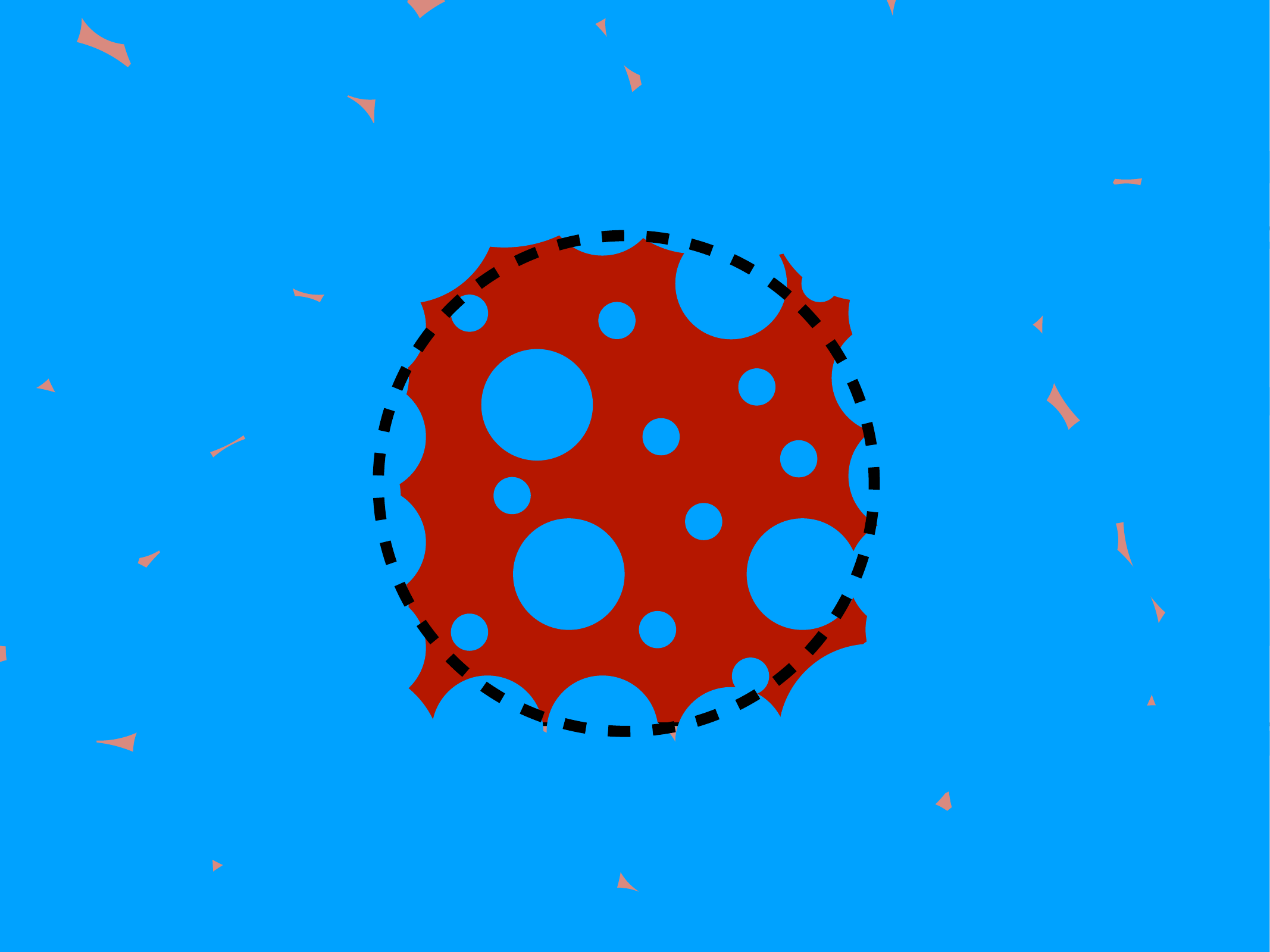}\\
    \caption{A schematic illustration for the delayed-decayed false vacuum regions to accumulate (false) vacuum energy densities that gradually dominate the evolution of these overdensities, slightly shrinking the local comoving Hubble horizon that effectively induces local close patches. The subsequent bubble nucleations within these local baby Universes would prevent them from entering into a state of eternal inflation and hence being detached from our Universe, ensuring the conditions for PBH formations.}
    \label{fig:schematic}
\end{figure*}

It is worth noting that a recent revisit~\cite{Flores:2024lng} of our accumulating mechanism has cast doubt on the PBH formations from the critical collapse criteria, but proposed to form PBHs from the Schwarzschild criterion for the growing energy density of the domain walls separating between the delayed-decayed and normal-decayed regions. They pointed out that, as the relative difference in the red-shifted radiation densities generates a local overdensity, the delayed-decayed regions are still dominated by radiation energy density, providing only expansionary solutions instead of a contracting solution as realized in the usual inflationary PBH scenarios that induce a local curvature from the curvature perturbations. Here we take the chance to clarify our accumulating mechanism that, the local overdensities are gained, not through the relative difference in the red-shifted radiation energy densities, but from the undiluted false vacuum energy density (fixing the true vacuum at zero energy) in the delayed-decayed regions. These false vacuum islands would gradually be dominated by the false vacuum energy and expand locally faster than the outside regions, inflating into a local baby Universe~\cite{Jinno:2023vnr} that can be treated effectively as a local closed Universe. Moreover, different from earlier proposals~\cite{Sato:1981bf,Maeda:1981gw,Sato:1981gv,Kodama:1981gu,Sato:1981hk,Kodama:1982sf} that might lead to type II PBHs (see, for example, a recent study~\cite{Uehara:2024yyp} and references herein), these delayed-decayed regions would not actually inflate into a detached baby Universe in the end due to subsequent bubble nucleations (see Fig.~\ref{fig:schematic}). Therefore, the original critical collapse criteria still apply in our case just as the usual inflationary PBH scenarios. 

Even if the postponed overdensities never reach the critical threshold for the PBH formations, they can still induce potentially observable super-horizon curvature perturbations~\cite{Liu:2022lvz} at small scales, whose null detection imposes extremely strong constraints on the very-strong super-cooling slow-type FOPTs at low-energy scales. The same strategy for estimating the curvature perturbations also applies to the inhomogeneous distributions of spherical domain walls~\cite{Zeng:2023jut} and the inhomogeneous catalyzations of FOPTs from PBHs~\cite{Zeng:2024snl}. An alternative evaluation was presented in a recent study~\cite{Elor:2023xbz} for the curvature perturbations sourced from the isocurvature perturbations in the dark sector inherited from super-horizon fluctuations in the completion time of some low-scale dark-sector FOPT. Similar treatment can be found in a recent study~\cite{Buckley:2024nen} for a high-scale FOPT. In particular, the distribution of overdensities at given scales from different transition patches was found in a recent study~\cite{Lewicki:2024ghw} to be non-Gaussian, which merits further studies as this would significantly impact the PBH mass and abundance as well as the power spectrum of curvature perturbations.

In this paper, we revisit and elaborate on our accumulating mechanism for producing PBHs and curvature perturbations in the delayed-decayed regions during a general cosmological FOPT set up in Sec.~\ref{sec:FOPT}. In Sec.~\ref{sec:island}, we solve for the evolution of these false vacuum islands by effectively transferring the inhomogeneities in the scale factors temporarily into the equation of state during the delayed-decayed process. In Sec.~\ref{sec:PBH}, we estimate numerically and fit analytically the PBH mass and abundance, but neglecting the effects from Refs.~\cite{Gouttenoire:2023naa,Lewicki:2023ioy} altogether for simplicity as the true-vacuum energy and bubble-wall energy propagating into the postponed Hubble patches from those bubbles nucleated inside the past light cone of collapsed regions have opposite effects on the growth of overdensities inside the postponed Hubble patches. In Sec.~\ref{sec:PR}, we directly compute the curvature perturbations by summing over all nucleation histories inside the postponed Hubble patches and truncating away the collapsed spacetime regions. The Sec.~\ref{sec:condis} is devoted to the conclusion and discussion for future perspective.

\section{General cosmological FOPTs}\label{sec:FOPT}

First, we set up the notations and conventions for a general cosmological FOPT. For a scalar field $\phi$ coupled to the background thermal plasma of temperature $T$ with an effective potential $V_\mathrm{eff}(\phi,T)$ admitting a false vacuum $V_+\equiv V_\mathrm{eff}(\phi_+)$ at $\phi_+$, there could develop a true vacuum $V_-(T)\equiv V_\mathrm{eff}(\phi_-(T),T)$ at $\phi_-(T)$ with decreasing temperature $T$. We omit the minor temperature dependence in $\phi_-(T)$ as usually done for simplicity. Suppose the scalar field $\phi$ initially populates at $\phi_+$ homogeneously over the space. In that case, the cosmological FOPT proceeds through random nucleations of true-vacuum bubbles in the false vacuum thermal background followed by rapid expansion until percolations via bubble collisions. The nucleation rate for a true-vacuum bubble per unit time and per unit volume is of an exponential form~\cite{Coleman:1977py,Callan:1977pt,Linde:1980tt,Linde:1981zj} $\Gamma(t)=A(t)e^{-B(t)}\equiv e^{-C(t)}$, where the pre-factor $A(t)$ accounts for quantum corrections of vacuum decay, and the bounce action $B(t)$ is evaluated from the Euclidean action at the bounce solution solved from the bounce equation given the boundary condition for describing a bubble of certain symmetry. In this paper, we study the usual case when $C(t)$ can be expanded linearly around some reference time $t_0$ as $C(t)=C(t_0)-\beta(t-t_0)$, in which case the nucleation rate can be approximated as $\Gamma(t)=e^{-C(t_0)+\beta(t-t_0)}=A(t_0)e^{-B(t_0)-\beta t_0}e^{\beta t}\equiv\Gamma_0e^{\beta t}$ with $\beta=\mathrm{d}\ln\Gamma/\mathrm{d}t|_{t=t_0}$. In what follows, we will use the mass scale $\Gamma_0^{1/4}$ to normalize all dimensional quantities to be dimensionless, for example, $\bar{\beta}\equiv\beta/\Gamma_0^{1/4}$ and $\bar{t}\equiv\Gamma_0^{1/4}t$. The above bubble nucleation rate $\Gamma(t_\mathrm{nuc})=H(t_\mathrm{nuc})^4$ defines a special moment $t_\mathrm{nuc}$ when balancing with the Hubble expansion rate $H(t_\mathrm{nuc})\equiv\dot{a}(t_\mathrm{nuc})/a(t_\mathrm{nuc})\equiv H_\mathrm{nuc}$, that is $\Gamma_0e^{\beta t_\mathrm{nuc}}=H_\mathrm{nuc}^4$, which leads to a would-be useful approximation $\beta/H_\mathrm{nuc}\approx8W(\bar{\beta}/8)$ with the Lambert function defined by $z=W(z)e^{W(z)}$ after approximating $H_\mathrm{nuc}t_\mathrm{nuc}\approx1/2$ within the radiation dominance.

Next, we introduce the probability of staying at the false vacuum, which is equivalent to the volume fraction of false vacuum, $F(t)=\overline{V}_f(t)/\overline{V}$, defined by the ratio of the comoving volume of false vacuum $\overline{V}_f(t)$ out of a total comoving volume $\overline{V}$. Thus, the decrease of false vacuum volume $\mathrm{d}\overline{V}_f(t)$ at $t$ due to the volume increase at $t$ from those bubbles nucleated at an earlier time $t'$ reads
\begin{align}
\mathrm{d}\overline{V}_f(t)=-\int_{t_i}^t\mathrm{d}N_b(t')\frac{\overline{V}_f(t)}{\overline{V}_f(t')}\mathrm{d}\overline{V}_b(t;t'),
\end{align}
where the number of bubbles nucleated at $t'$ within a time interval $\mathrm{d}t'$ is $\mathrm{d}N_b(t')=\Gamma(t')V_f(t')\mathrm{d}t'$ by definition as bubbles can only be nucleated within the physical false vacuum volume $V_f(t')=\overline{V}_f(t')a(t')^3$ with a scale factor $a(t')$, and the ratio $\overline{V}_f(t)/\overline{V}_f(t')$ simply takes into account the fact that only the parts of bubbles propagating into the false vacuum phase can decrease the false vacuum fraction~\cite{Enqvist:1991xw,Hindmarsh:2019phv}. The increasing volume $\mathrm{d}\overline{V}_b(t;t')$ from true-vacuum bubbles at $t$ nucleated at an earlier time $t'$ vanishes if $t=t'$. Therefore, the false vacuum fraction evolves according to the equation
\begin{align}
\frac{\mathrm{d}F(t)}{\mathrm{d}t}=F(t)\left[-\int_{t_i}^t\mathrm{d}t'\Gamma(t')a(t')^3\frac{\mathrm{d}\overline{V}_b(t;t')}{\mathrm{d}t}\right],
\end{align}
which can be directly solved as~\cite{Guth:1982pn,Turner:1992tz}
\begin{align}\label{eq:Foft}
F(t;t_i)=\exp\left[-\int_{t_i}^t\mathrm{d}t'\Gamma(t')a(t')^3\overline{V}_b(t;t')\right].
\end{align}
Here $t_i$ is the earliest moment when the bounce equation of $\phi$ ever allows for an instanton solution, and hence all regions are in the false vacuum before $t_i$, that is $F(t<t_i; t_i)=1$. Due to the friction term in the bounce equation, $t_i$ is slightly later than the critical time $t_c$ when $V_+$ and $V_-$ are degenerated. The comoving radius $r(t;t')$ in computing the comoving volume $\overline{V}_b(t;t')=\frac43\pi r(t;t')^3$ of a true-vacuum bubble at $t$ nucleated at an earlier time $t'$ can be estimated as
\begin{align}
r(t;t')\equiv\frac{r_0}{a(t')}+\int_{t'}^t\frac{v_w(\tilde{t})\mathrm{d}\tilde{t}}{a(\tilde{t})}\approx (v_w\equiv1)\int_{t'}^t\frac{\mathrm{d}\tilde{t}}{a(\tilde{t})},
\end{align}
where the initial bubble radius $r_0$ is usually ignored compared to its following expansion, and the full-time evolution of the bubble wall velocity $v_w(t)$~\cite{Cai:2020djd} is also ignored by directly taking its terminal value $v_w$ to be unity for simplicity (the other constant value is a direct generalization, see, for example, Ref.~\cite{He:2023ado}). The above false vacuum fraction function also defines a special moment $t_\mathrm{per}$ when most of the bubbles start to percolate with $F(t_\mathrm{per})=0.7$~\cite{percolation1971}, roughly coinciding with the time $t_*$ when the associated GW background peaks in its energy spectrum with maximal strength~\cite{Cai:2017tmh}. For any FOPT, it usually holds $t_c<t_i<t_\mathrm{nuc}<t_\mathrm{per}<t_*$. We will identify $t_*=t_\mathrm{per}$ hereafter.

Finally, we depict the general picture of a cosmological FOPT inside and outside the delayed-decayed regions. Before the initial time for bubble nucleations, the Universe is dominated by radiation energy density diluted as $\rho_r(t<t_i)\propto a(t)^{-4}$ with an expanding scale factor $a(t)$, while the false vacuum energy density stays as a sub-dominated constant, $\rho_v(t<t_i)\equiv V_+<\rho_r(t_i)$. Ever since the first nucleation of a bubble after the time $t_i$, the Universe is still globally dominated by radiations, but locally the delayed-decayed false vacuum regions are gradually dominated by the vacuum energy densities, and hence the scale factor would not be homogeneous anymore with more bubbles nucleated and percolated after rapid expansion and violent collisions. In both false vacuum and true vacuum regions, the radiations are still red-shifted as $\rho_r(t>t_i, \mathbf{x}_+|\phi(t, \mathbf{x}_+)=\phi_+)\propto a(t, \mathbf{x}_+)^{-4}$ and $\rho_r(t>t_i, \mathbf{x}_-|\phi(t, \mathbf{x}_-)=\phi_-)\propto a(t, \mathbf{x}_-)^{-4}$, respectively, but with different scale factors, while the vacuum energy densities stay at different constants $\rho_v(t>t_i, \mathbf{x}_+)\equiv V_+>V_-\equiv\rho_v(t>t_i, \mathbf{x}_-)$. It is this vacuum-energy density difference $\Delta V\equiv V_+-V_-$ that accumulates over time in false vacuum regions compared to true-vacuum regions with respect to rapidly diluted radiation energy densities. Hence, the evolution equations read
\begin{align}\label{eq:EOM1}
\frac{\mathrm{d}}{\mathrm{d}t}\rho_r(t,\mathbf{x}_\pm)+4H(t,\mathbf{x}_\pm)\rho_r(t,\mathbf{x}_\pm)=-\frac{\mathrm{d}}{\mathrm{d}t}\rho_v(t,\mathbf{x}_\pm),
\end{align}
\begin{align}\label{eq:EOM2}
3M_\mathrm{Pl}^2H(t,\mathbf{x}_\pm)^2=\rho_r(t,\mathbf{x}_\pm)+\rho_v(t,\mathbf{x}_\pm),
\end{align}
where the spatial dependence $\mathbf{x}_\pm$ can be attributed to different nucleation time of the first bubble locally in that patch with respect to the earliest nucleation time $t_i$ globally, that is, $\rho_r(t,\mathbf{x}_\pm)=\rho_r(t;t_i+\Delta t(\mathbf{x}_\pm))\equiv\rho_r(t;t_n(\mathbf{x}_\pm))$, correspondingly solved from $\rho_v(t,\mathbf{x}_+)\equiv\rho_v(t;t_n(\mathbf{x}_+))=F(t;t_n)V_+$ and $\rho_v(t,\mathbf{x}_-)\equiv\rho_v(t;t_n(\mathbf{x}_-))=[1-F(t;t_n)]V_-=0$ after setting the zero of the potential at $V_-$. Note here that the energy densities of expanding bubble walls have already been included in the total radiations as seen from directly estimating the equation of state~\cite{Liu:2021svg,Liu:2022lvz},
\begin{align}
w_\mathrm{wall}=\frac{p_\mathrm{wall}}{\rho_\mathrm{wall}}=\frac{\int\mathrm{d}z\left[\frac12\left(\frac{\mathrm{d}\phi}{\mathrm{d}t}\right)^2-\frac{1}{6a^2}\left(\frac{\mathrm{d}\phi}{\mathrm{d}z}\right)\right]}{\int\mathrm{d}z\left[\frac12\left(\frac{\mathrm{d}\phi}{\mathrm{d}t}\right)^2+\frac{1}{2a^2}\left(\frac{\mathrm{d}\phi}{\mathrm{d}z}\right)\right]}\approx\frac13,
\end{align}
where we have used the approximation $\mathrm{d}\phi/\mathrm{d}t\approx\mathrm{d}\phi/\mathrm{d}z/a$ for the bubble wall velocity close to the speed of light. Even when bubble walls collide, they are still red-shifted as the dissipated thermal radiations and associated GW radiations. This evolutionary scaling for bubble walls is also consistent with the recent estimations~\cite{Gouttenoire:2023naa,Lewicki:2023ioy}.

\section{False vacuum islands}\label{sec:island}

\begin{figure*}
    \centering
    \includegraphics[width=0.49\textwidth]{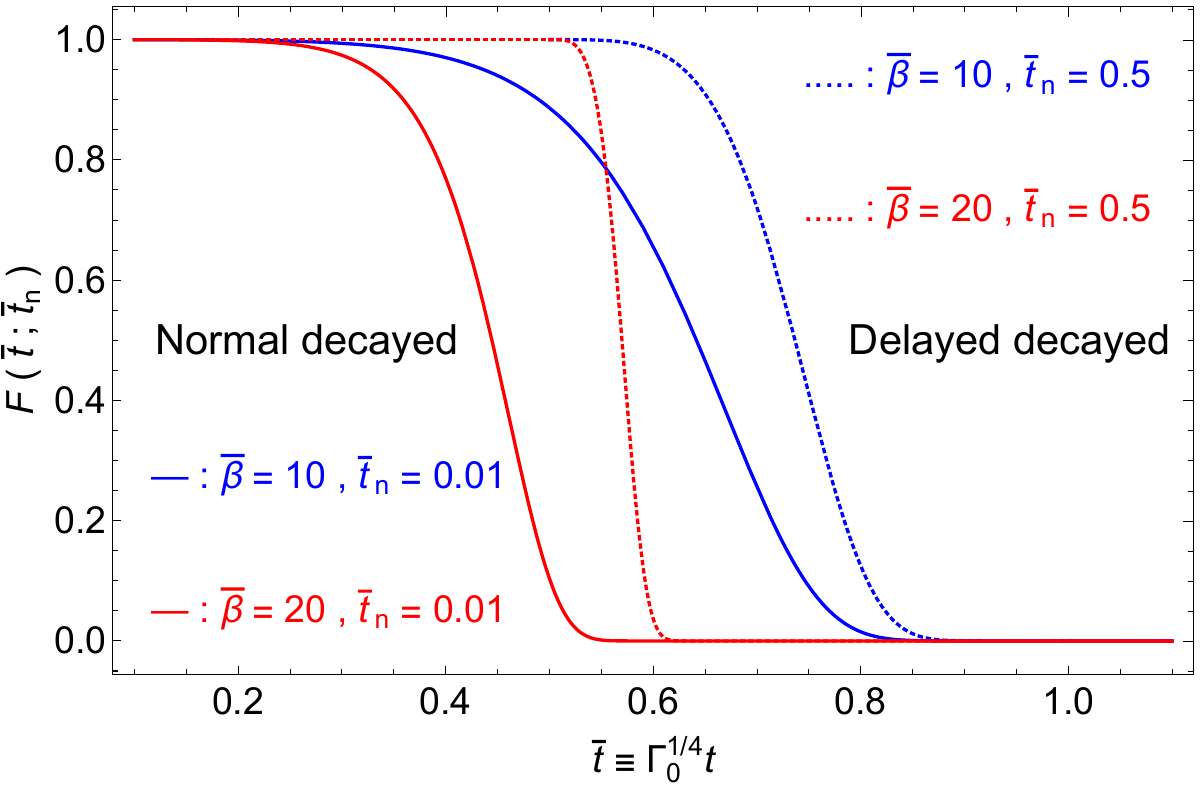}
    \includegraphics[width=0.47\textwidth]{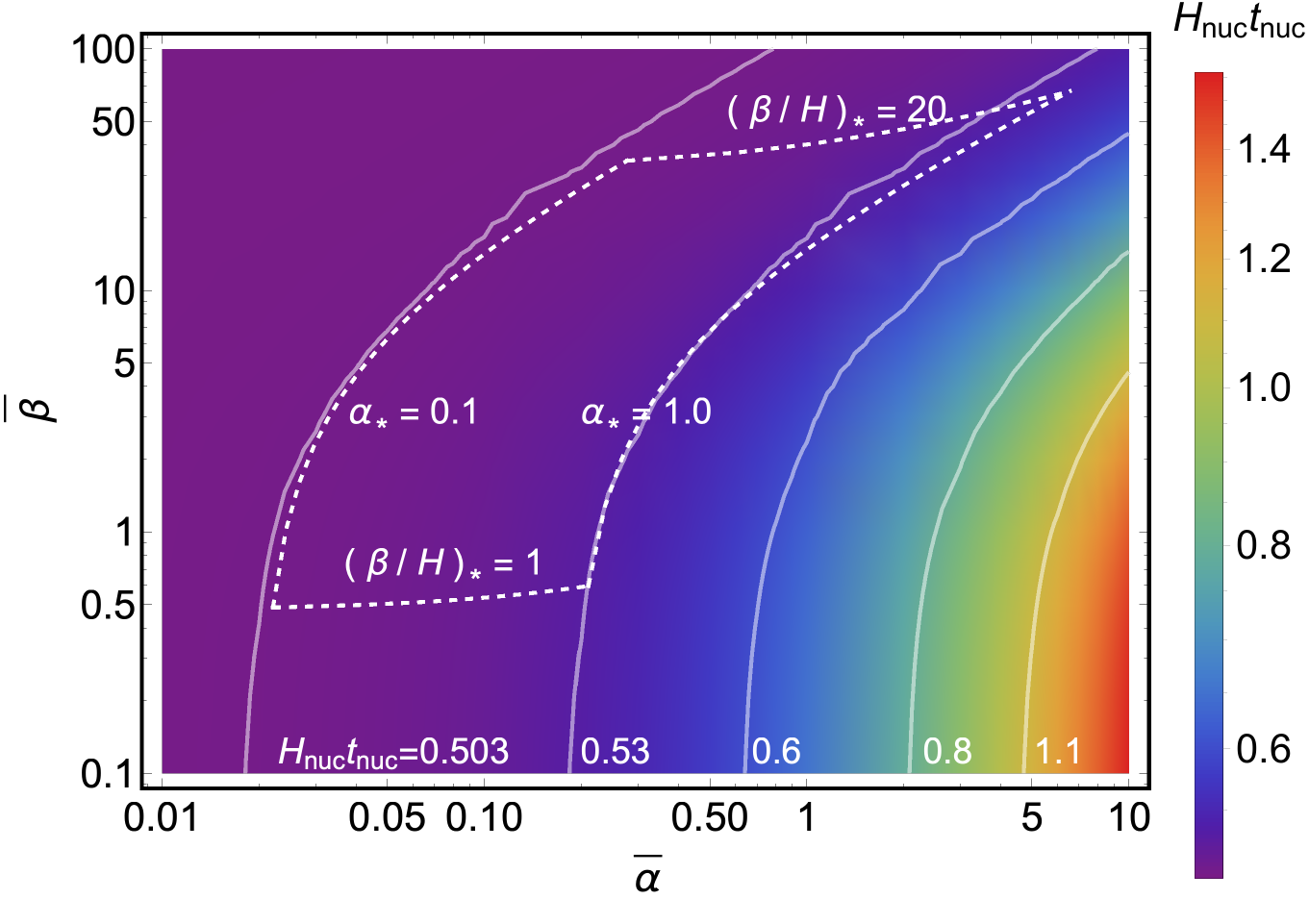}\\
    \caption{\textit{Left}: The time evolution of the false vacuum fraction given earlier (solid) and later (dashed) nucleation times of the first bubble with shorter (red) and longer (blue) duration times. \textit{Right}: The physical (phenomenological) strength and duration parameters are bounded in the region with $0.1<\alpha_*<1.0$ and $1<(\beta/H)_*<20$ (dashed) given the model parameters $\bar{\alpha}$ and $\bar{\beta}$. The approximation $H_\mathrm{nuc}t_\mathrm{nuc}\approx1/2$ is checked with solid contour curves.}
    \label{fig:FOPT}
\end{figure*}

\begin{figure*}
    \centering
    \includegraphics[width=0.48\textwidth]{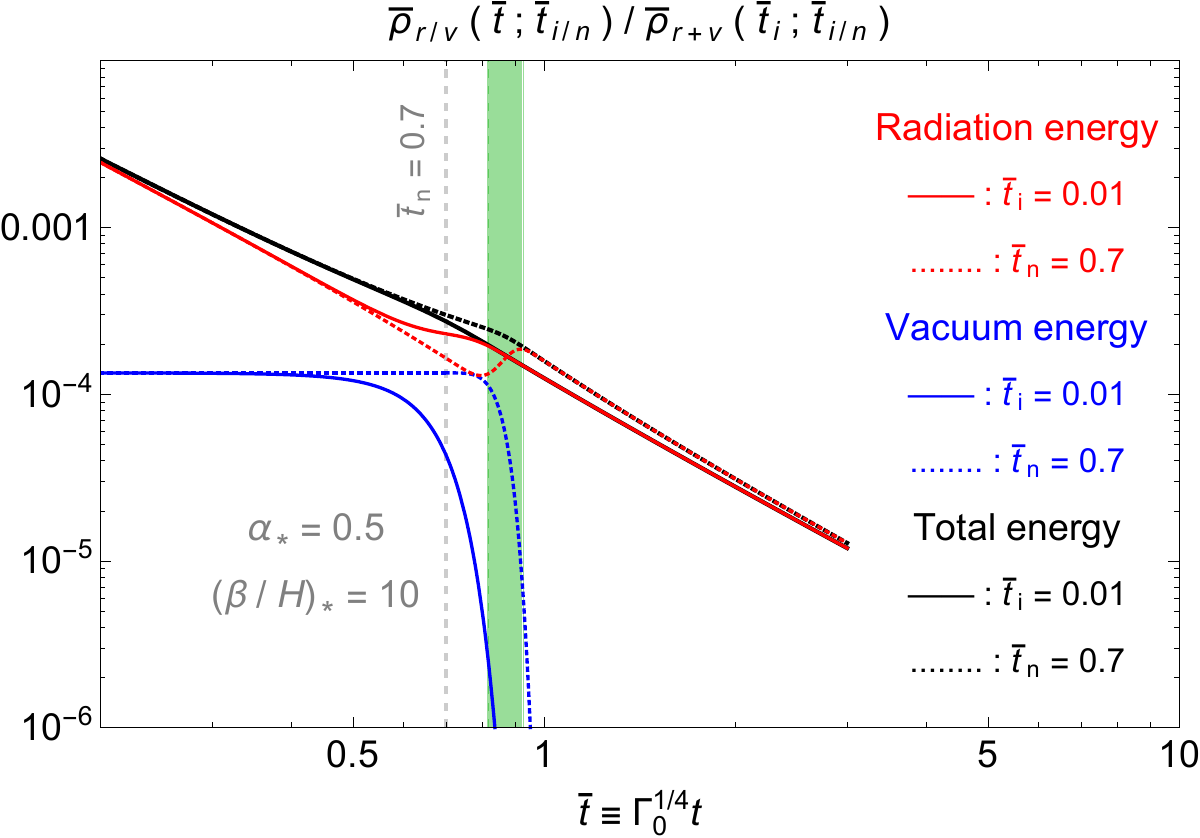}
    \includegraphics[width=0.48\textwidth]{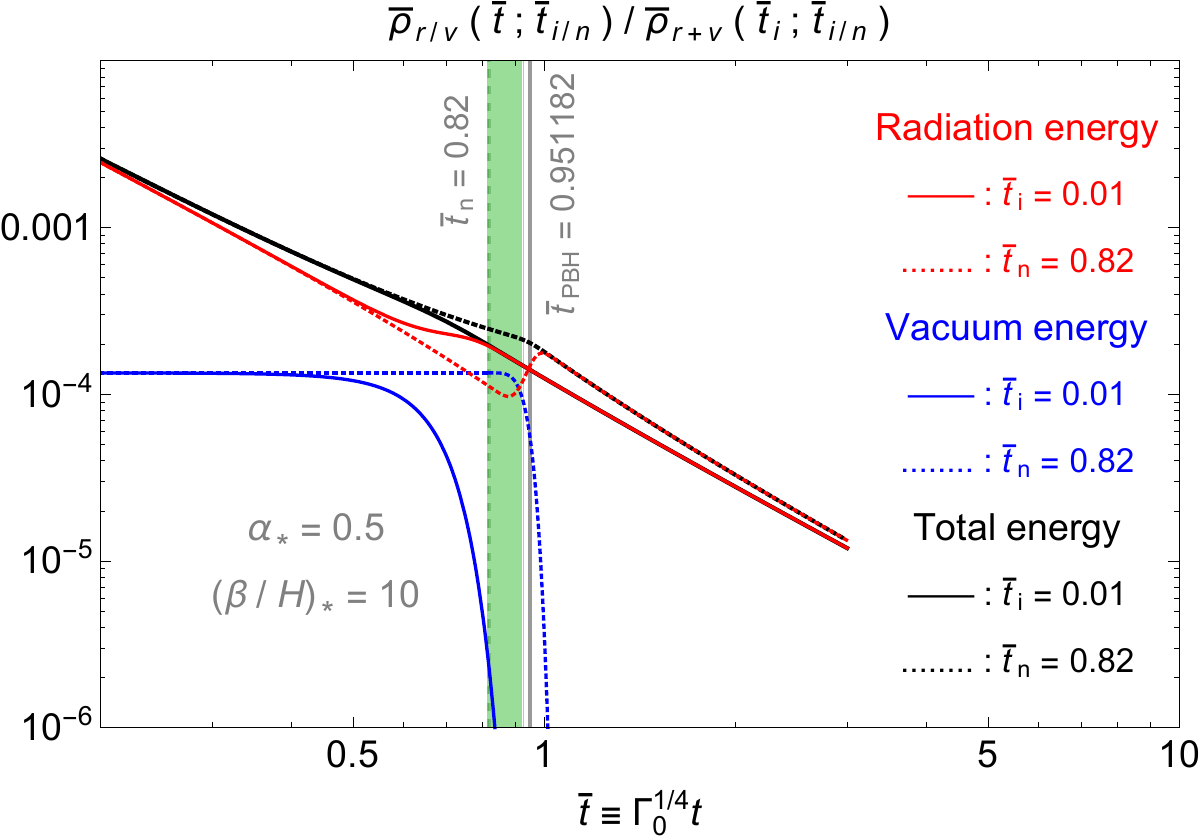}\\
    \includegraphics[width=0.48\textwidth]{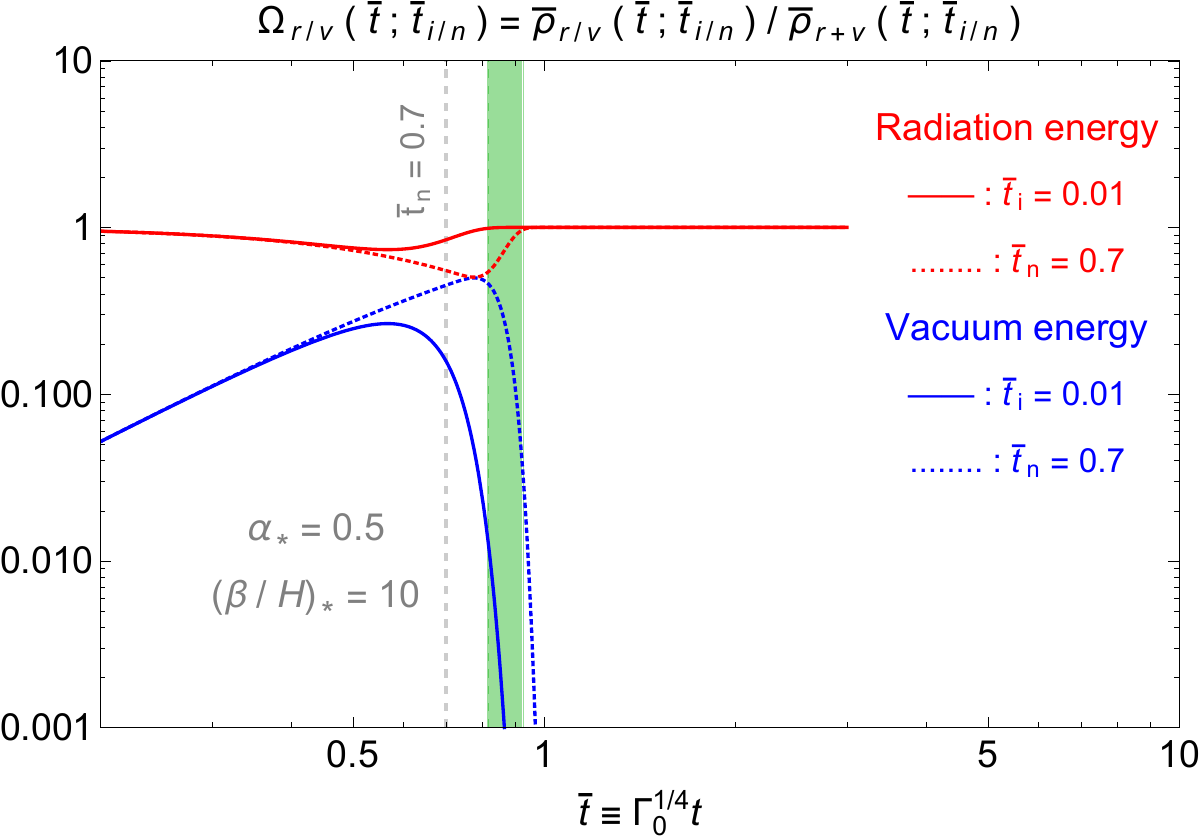}
    \includegraphics[width=0.48\textwidth]{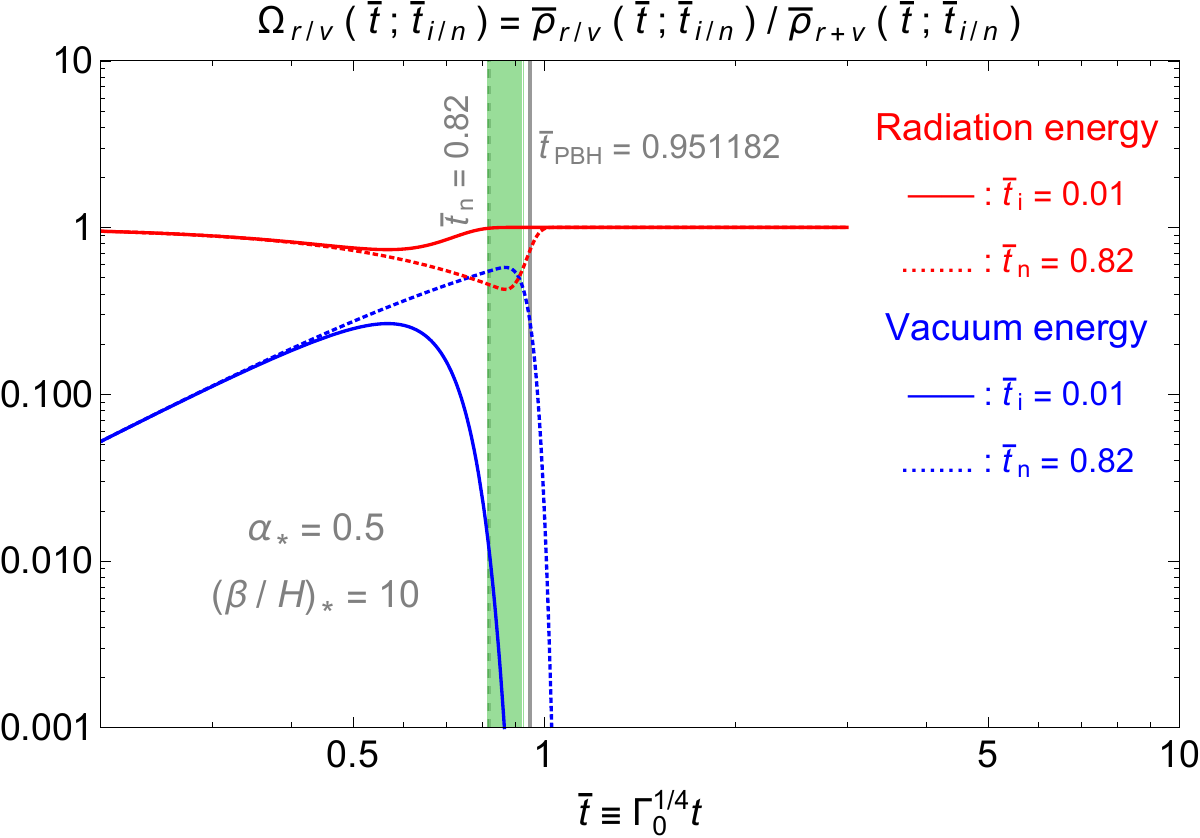}\\
    \caption{The time evolutions of energy densities with respect to the total density at the initial nucleation time (top) and the real time (bottom) for radiations (red) and vacuum (blue) components as well as their combined (black) in the normal-decayed (solid) and delayed-decayed (dashed) regions during an illustrative FOPT with a strength parameter $\alpha_*=0.5$ and a duration parameter $(\beta/H)_*=10$. PBHs can be formed at the latest at the gray vertical line (right) for a delayed-decayed nucleation time at the earliest at the left boundary of the green band, below which no PBH can be produced (left).}
    \label{fig:solutioins}
\end{figure*}

Solving the combined equations~\eqref{eq:Foft},~\eqref{eq:EOM1}, and~\eqref{eq:EOM2} is inefficient as they are coupled integro-differential equations of the scale factors in inhomogeneous regions. One rigorous method is to solve Eqs.~\eqref{eq:EOM1} and~\eqref{eq:EOM2} iteratively with some initial test form for the scale factor in the false vacuum fraction~\eqref{eq:Foft} as done recently in Ref.~\cite{Kanemura:2024pae}, which might still be inefficient for our later purpose to calculate the curvature perturbations. The other approach is to transform the two-dimensional system of integro-differential equations into a seven-dimensional system of first-order ordinary differential equations as proposed recently in Ref,~\cite{Flores:2024lng}, which we will pursue in future studies. In this study, we will adopt an approximated strategy. For a general FOPT that can be completed appropriately~\cite{Guth:1982pn,Turner:1992tz,Ellis:2018mja}, the global evolutions before and after the FOPT are both dominated by radiations diluted as $\rho_r(t)\propto a(t)^{-4}$ with the scale factor $a(t)\propto t^{1/2}$. However, the radiation energy densities right before and after the FOPT will differ by the released vacuum energy densities that turn into radiation-like walls, and hence there is a step-like jumping in the time evolution of total radiations, during which the radiations are always red-shifted as $\rho_r\propto a^{-4}$ but the scale factor would expand faster than $a(t)\propto t^{1/2}$. The approximated strategy we adopt is to effectively transfer the inhomogeneities in the scale factors into the equation of state of inhomogeneous distributions of radiations. This is done by first fixing the scale factor in the false vacuum fraction~\eqref{eq:Foft} as in the radiation-dominated era with $a(t)\propto t^{1/2}$ throughout the FOPT as shown in the left panel of  Fig.~\ref{fig:FOPT}, 
\begin{align}\label{eq:FoftRD}
F(\bar{t};\bar{t}_n)=\exp\left[-\frac43\pi\Theta(\bar{t}-\bar{t}_n)\int_{\bar{t}_n}^{\bar{t}}\mathrm{d}\bar{t}'e^{\bar{\beta}\bar{t}'}8\left(\sqrt{\bar{t}\bar{t}'}-\bar{t}'\right)^3\right],
\end{align}
and then solve the dimensionless version of the dynamical equations~\eqref{eq:EOM1} and~\eqref{eq:EOM2},
\begin{align}\label{eq:EOMRD}
\frac{\mathrm{d}}{\mathrm{d}\bar{t}}\bar{\rho}_r(\bar{t};\bar{t}_n)+4\sqrt{\bar{\rho}_r(\bar{t};\bar{t}_n)+\bar{\alpha}F(\bar{t};\bar{t}_n)}\bar{\rho}_r(\bar{t};\bar{t}_n)=-\bar{\alpha}\dot{F}(\bar{t};\bar{t}_n),
\end{align}
with an initial condition deep into the radiation era,
\begin{align}\label{eq:ICRD}
\bar{\rho}_r(\bar{t}_\mathrm{ini};\bar{t}_n)=\frac{1}{4\bar{t}_\mathrm{ini}^{\,2}}-\bar{\alpha},
\end{align}
where the dimensionless radiation density $\bar{\rho}_r(\bar{t};\bar{t}_n)\equiv\rho_r(\bar{t};\bar{t}_n)/(3M_\mathrm{Pl}^2\Gamma_0^{1/2})$ is defined with respect to a characteristic density scale $\rho_\Gamma\equiv3M_\mathrm{Pl}^2\Gamma_0^{1/2}\equiv3M_\mathrm{Pl}^2H_\Gamma^2$ that can be used to normalize a dimensionless strength factor $\bar{\alpha}\equiv\Delta V/\rho_\Gamma$. Thus, the initial condition is obtained by 
\begin{align}
&\bar{\rho}_r(\bar{t}_\mathrm{ini};\bar{t}_n)\equiv\frac{\rho_r(\bar{t}_\mathrm{ini};\bar{t}_n)}{3M_\mathrm{Pl}^2H_\Gamma^2}=\frac{3M_\mathrm{Pl}^2H_\mathrm{ini}^2}{3M_\mathrm{Pl}^2H_\Gamma^2}-\frac{\rho_v(\bar{t}_\mathrm{ini};\bar{t}_n)}{\rho_\Gamma}\nonumber\\
&=\left(\frac{H_\mathrm{ini}t_\mathrm{ini}}{\Gamma_0^{1/4}t_\mathrm{ini}}\right)^2-\frac{F(\bar{t}_\mathrm{ini};\bar{t}_n)\Delta V}{\rho_\Gamma}=\frac{1}{4\bar{t}_\mathrm{ini}^{\,2}}-\bar{\alpha},
\end{align}
where we have used $H_\mathrm{ini}t_\mathrm{ini}=1/2$ for any initial time $t_\mathrm{ini}\ll t_i$ sufficiently deep into the radiation dominance without nucleating any bubble yet with $F(\bar{t}_\mathrm{ini};\bar{t}_n)=1$.

Therefore, the normal decayed regions can be solved from~\eqref{eq:FoftRD}, \eqref{eq:EOMRD}, and~\eqref{eq:ICRD} by fixing the nucleation time of the first bubble at the earliest nucleation time globally, that is $t_n=t_i$, while the delayed-decayed regions can be solved from ~\eqref{eq:FoftRD}, \eqref{eq:EOMRD}, and~\eqref{eq:ICRD} by choosing any nucleation time $t_n>t_i$ later than the global one. The global initial nucleation time $\bar{t}_i$ can be arbitrarily fixed as long as it meets $\bar{t}_i<\bar{t}_\mathrm{nuc}\equiv\Gamma_0^{1/4}t_\mathrm{nuc}=H_\mathrm{nuc}t_\mathrm{nuc}e^{-\beta/(8H_\mathrm{nuc})}\approx(1/2)e^{-W(\bar{\beta}/8)}=(4/\bar{\beta})W(\bar{\beta}/8)$ from recalling that $\Gamma_0^{1/4}\equiv H_\mathrm{nuc}e^{-\beta/(8H_\mathrm{nuc})}$, $H_\mathrm{nuc}t_\mathrm{nuc}\approx 1/2$, and $\beta/(8H_\mathrm{nuc})\approx W(\bar{\beta}/8)$. It is worth noting that, both the model parameters $\bar{\alpha}\equiv\Delta V/(3M_\mathrm{Pl}^2\Gamma_0^{1/2})$ and $\bar{\beta}\equiv\beta/\Gamma_0^{1/4}$ are not equal to the physical strength and duration parameters $\alpha_*$ and $(\beta/H)_*$ defined phenomenologically for general cosmological FOPTs by
\begin{align}\label{eq:parameters}
\alpha_*&=\frac{\Delta V}{\rho_r(t_\mathrm{per};t_i)}=\frac{\bar{\alpha}}{\bar{\rho}_r(\bar{t}_\mathrm{per};\bar{t}_i)},\\
\left(\frac{\beta}{H}\right)_*&=\frac{\beta/\Gamma_0^{1/4}}{\sqrt{H_*^2/\Gamma_0^{1/2}}}=\frac{\bar{\beta}}{\sqrt{\bar{\rho}_\mathrm{tot}(\bar{t}_\mathrm{per};\bar{t}_i)}},
\end{align}
where the total dimensionless density is generally defined as  $\bar{\rho}_\mathrm{tot}(\bar{t};\bar{t}_n)\equiv\bar{\rho}_r(\bar{t};\bar{t}_n)+\bar{\rho}_v(\bar{t};\bar{t}_n)=\bar{\rho}_r(\bar{t};\bar{t}_n)+\bar{\alpha}F(\bar{t};\bar{t}_n)$. In this paper, we will illustrative with a general FOPT in the physical parameter space bounded by $0.1<\alpha_*<1.0$ and $1<(\beta/H)_*<20$, which can be mapped onto the model parameter space in the $\bar{\alpha}-\bar{\beta}$ plane as shown in the right panel of Fig.~\ref{fig:FOPT} with the approximated relation $H_\mathrm{nuc}t_\mathrm{nuc}\approx1/2$ checked by the colored contoured regions. In this parameter region, we can safely choose the global initial nucleation time at $\bar{t}_i=0.01$, any other values comparable or earlier than this one are also allowable.

\section{Primordial black holes}\label{sec:PBH}

\begin{figure*}
    \centering
    \includegraphics[width=0.48\textwidth]{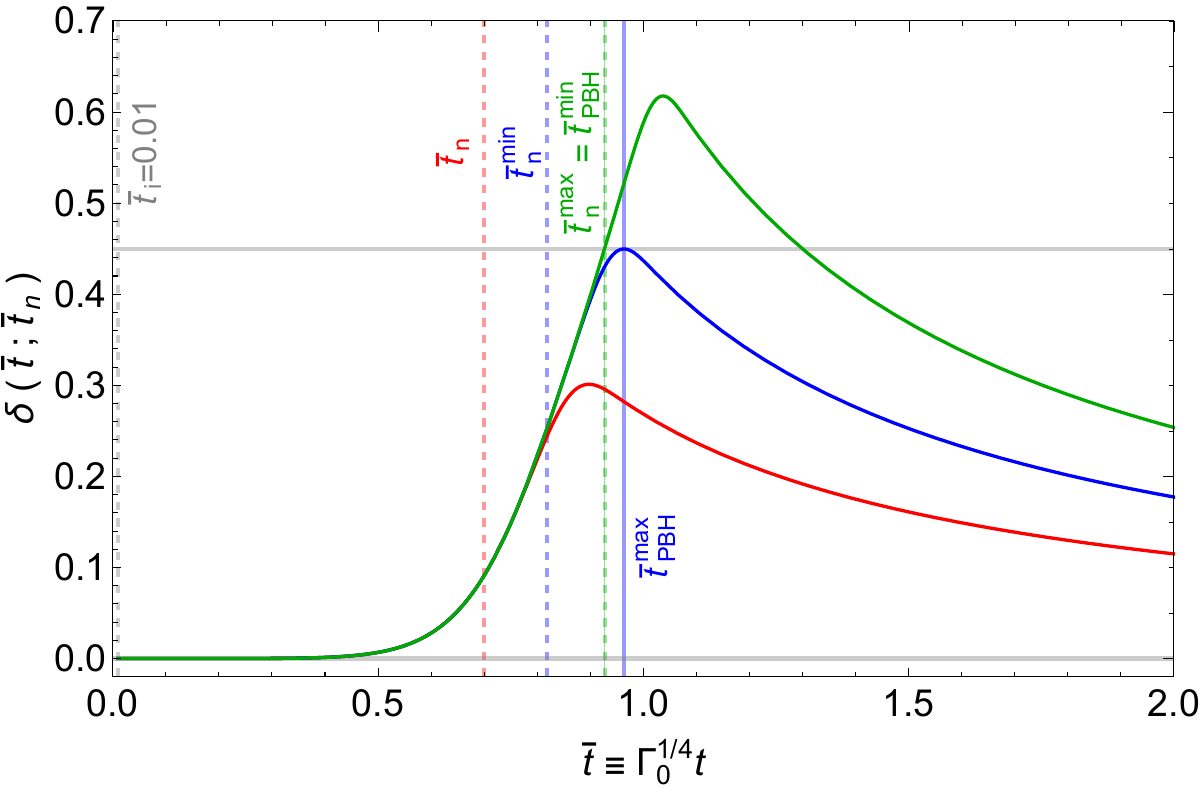}
    \includegraphics[width=0.48\textwidth]{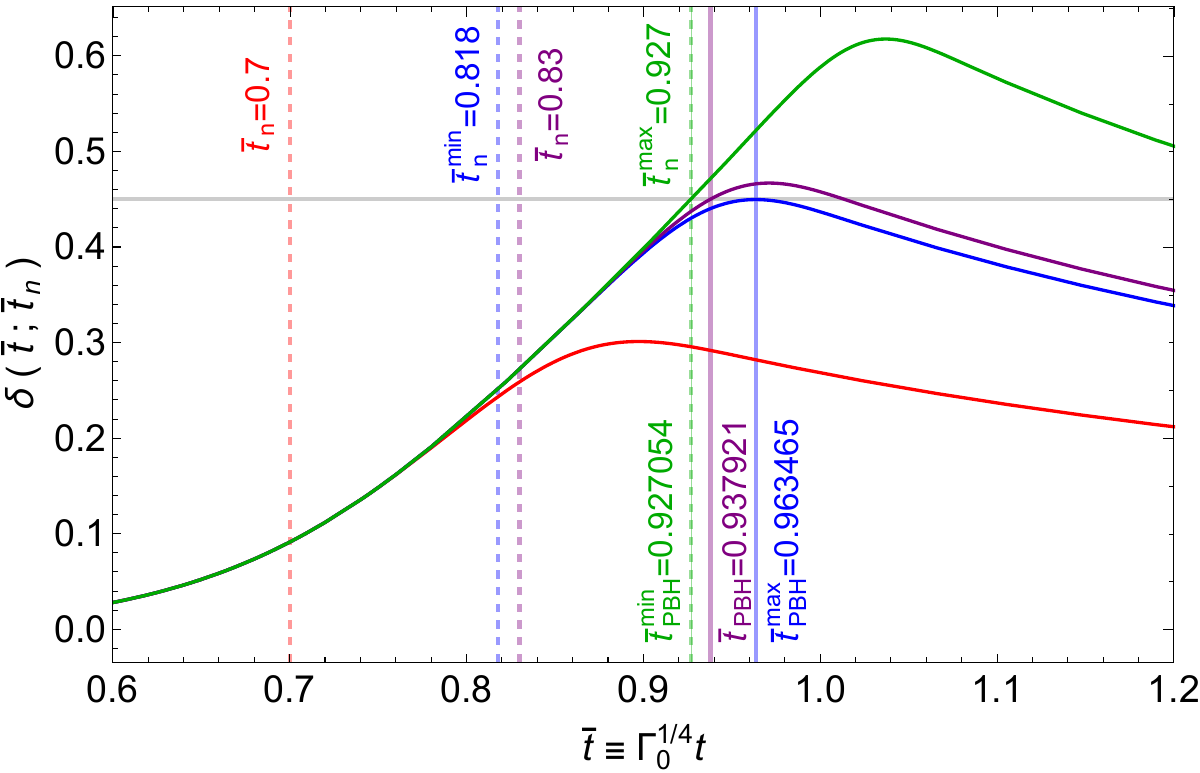}\\
    \caption{The time evolution of the overdensity (left) and its zoom-in view (right) in the delayed-decayed region during a FOPT with illustrative strength factor $\alpha_*=0.5$ and inverse duration $(\beta/H)_*=10$, where vertical dashed lines are different delayed-decayed nucleation times of the first bubble with possible PBH formation at later times as shown with vertical solid lines.}
    \label{fig:overdensities}
\end{figure*}

Given the model parameters $\bar{\alpha}$ and $\bar{\beta}$, both the radiation energy density $\bar{\rho}_r(\bar{t};\bar{t}_n)$ and (false) vacuum energy density $\bar{\rho}_v(\bar{t};\bar{t}_n)\equiv\bar{\alpha}F(\bar{t};\bar{t}_n)$ can be directly solved in the normal-decayed and delayed-decayed regions with $\bar{t}_n=\bar{t}_i$ and $\bar{t}_n>\bar{t}_i$, respectively. Note that one can use these solutions to directly compute the physical strength and duration parameters via~\eqref{eq:parameters}, which can also be used inversely to determine the model parameters $\bar{\alpha}$ and $\bar{\beta}$ in terms of the physical parameters $\alpha_*$ and $(\beta/H)_*$. The time evolutions of these solutions are presented in Fig.~\ref{fig:solutioins} for an illustrative FOPT with physical strength factor $\alpha_*=0.5$ and inverse duration $(\beta/H)_*=10$ with (right) and without (left) the outcome of PBH formations. To see when PBHs can form, we first define the energy density contrast in the delayed-decayed regions with respect to the normal-decayed regions as
\begin{align}
\delta(\bar{t}; \bar{t}_n, \bar{t}_i)=\frac{\bar{\rho}_r(\bar{t};\bar{t}_n)+\bar{\rho}_v(\bar{t};\bar{t}_n)}{\bar{\rho}_r(\bar{t};\bar{t}_i)+\bar{\rho}_v(\bar{t};\bar{t}_i)}-1,
\end{align}
where we choose the normal-decayed regions in the denominator with the earliest possible nucleation time of the first bubble globally. This is justified as the probability of staying in the false vacuum to accumulate (false) vacuum energy density is exponentially suppressed by another exponential approximately. That is to say, a region with some local nucleation time is more probably surrounded by the regions with earlier nucleation times, and hence the local energy density contrast always tends to be over-dense rather than under-dense, which can be effectively achieved by choosing the nucleation time in the denominator as the global one.

Next, we turn to the overdensity evolution as shown in Fig.~\ref{fig:overdensities}, where the right panel is just a zoom-in view of the left panel, and all vertical dashed lines are the corresponding local nucleation times with possible PBH formation times as shown with vertical solid lines. For a longer enough delayed decay time $\bar{t}_n$ as shown with the purple curve in the right panel, it is always possible to locate a time $\bar{t}_\mathrm{PBH}$ for surpassing the PBH threshold with $\delta(\bar{t}_\mathrm{PBH}; \bar{t}_n, \bar{t}_i)=\delta_c$ at the vertical purple line in the right panel. However, there is a lower bound on the delayed decay time $\bar{t}_n^\mathrm{min}$ as shown with a vertical blue dashed line, below which the overdensity never reaches the PBH threshold $\delta_c=0.45$~\cite{Musco:2004ak,Harada:2013epa} as shown with the red curve. That is to say, for a delayed-decayed region with this local nucleation time $\bar{t}_n^\mathrm{min}$ as shown with the blue curve, the overdensity would saturate the PBH threshold exactly at the maximum of the overdensity evolution,
\begin{align}\label{eq:MostPBH}
\delta(\bar{t}_\mathrm{PBH}^\mathrm{max}; \bar{t}_n^\mathrm{min},\bar{t}_i)&=\delta_c,\\
\frac{\mathrm{d}}{\mathrm{d}\bar{t}}\delta(\bar{t}; \bar{t}_n^\mathrm{min}, \bar{t}_i)\bigg|_{\bar{t}=\bar{t}_\mathrm{PBH}^\mathrm{max}}&=0,\\
\frac{\mathrm{d}^2}{\mathrm{d}\bar{t}^2}\delta(\bar{t}; \bar{t}_n^\mathrm{min}, \bar{t}_i)\bigg|_{\bar{t}=\bar{t}_\mathrm{PBH}^\mathrm{max}}&<0.
\end{align}
Here the maximum point is denoted by $\bar{t}_\mathrm{PBH}^\mathrm{max}$ as this is the latest time for PBH formations as shown with the vertical blue solid line. To see this, note that regions with earlier nucleation time (for example, the blue curve) will accumulate the overdensity slower than regions with later nucleation time (for example, the purple curve), and hence will form PBHs at a later time as shown with the vertical blue solid line compared to the vertical purple solid line. Therefore, the earliest possible nucleation time $\bar{t}_n^\mathrm{min}$ would correspond to the latest possible PBH formation time $\bar{t}_\mathrm{PBH}^\mathrm{max}$. Similarly, there is also an upper bound on the delayed decay nucleation time $\bar{t}_n^\mathrm{max}$ as shown with the vertical green dashed line, in which case the local nucleation time $\bar{t}_n$ is late enough so that the corresponding PBH formation time $\bar{t}_\mathrm{PBH}$ becomes early enough to exactly coincide with the local nucleation time so that the PBH threshold $\delta(\bar{t}_\mathrm{PBH}^\mathrm{min}; \bar{t}_n^\mathrm{max},\bar{t}_i)=\delta_c$ is exactly found when $\bar{t}_n^\mathrm{max}=\bar{t}_\mathrm{PBH}^\mathrm{min}$ as shown with the vertical green dashed/solid lines. Except for this critical case, all delayed-decayed regions have already started the process of bubble nucleation before PBH formations as re-stressed recently in Ref.~\cite{Gouttenoire:2023naa}.

\begin{figure*}
    \centering
    \includegraphics[width=0.48\textwidth]{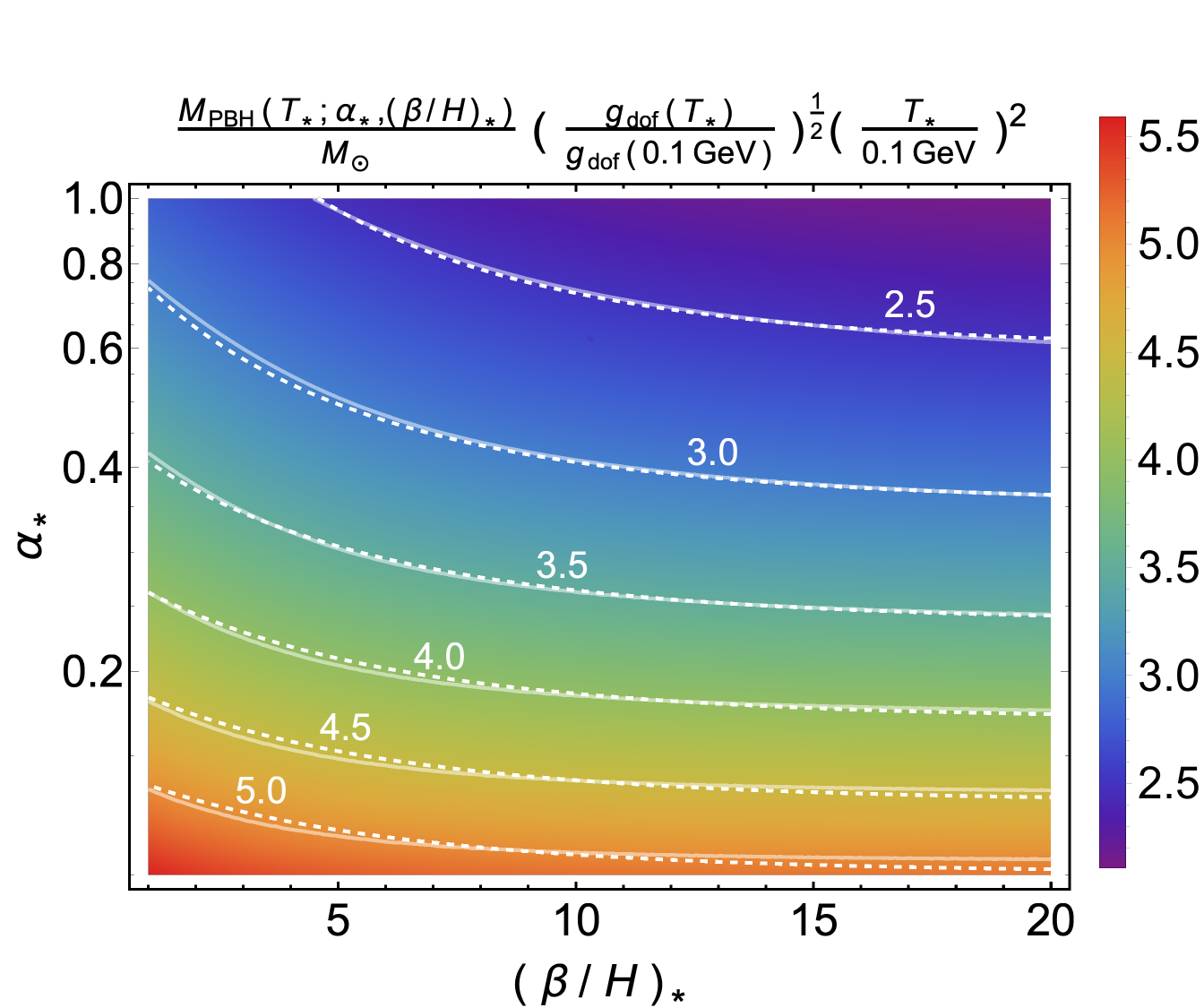}
    \includegraphics[width=0.48\textwidth]{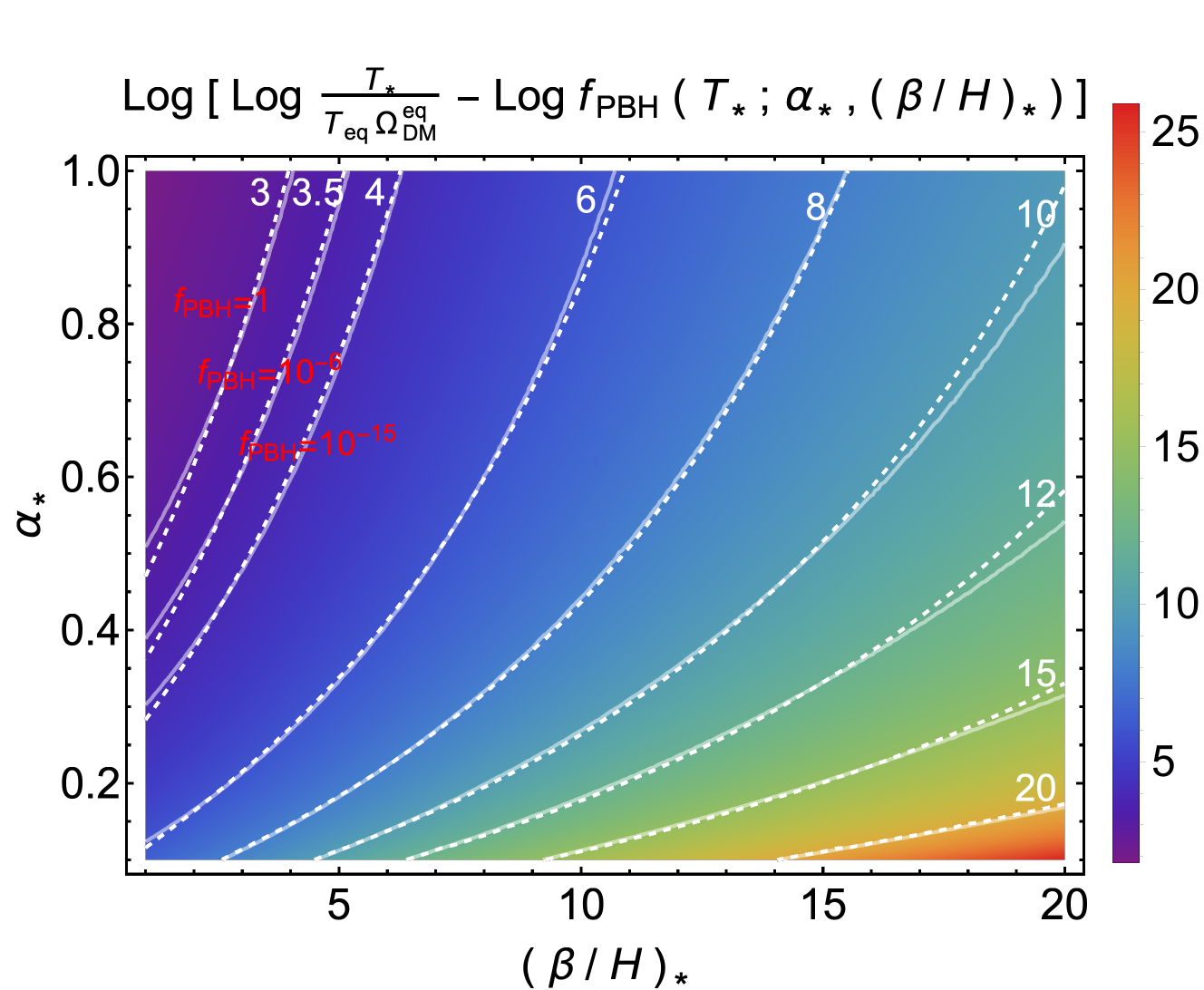}\\
    \caption{The fitting formulas (dashed) to the PBH mass (left) and abundance (right) given the FOPT parameters from the strength factor $\alpha_*$, inverse duration $(\beta/H)_*$, and percolation temperature $T_*$ for the most probable case of PBH formations. }
    \label{fig:FitPBH}
\end{figure*}

Finally, we determine that the most probable time for PBH formations is $\bar{t}_\mathrm{PBH}^\mathrm{max}$ since it corresponds to the earliest delayed decay nucleation time $\bar{t}_n^\mathrm{min}$, after which the probability of staying at the false vacuum for such a long time is exponentially suppressed by another exponential, and hence the corresponding PBH abundance must be negligible. Therefore, the PBH mass function $f_\mathrm{PBH}(M_\mathrm{PBH})$ for our accumulating mechanism is almost monochromatic. For the sake of most generality, we will estimate the PBH mass function for arbitrary $t_n^\mathrm{min}\leq\bar{t}_n\leq\bar{t}_n^\mathrm{max}$ and corresponding $\bar{t}_\mathrm{PBH}^\mathrm{min}\leq\bar{t}_\mathrm{PBH}\leq\bar{t}_\mathrm{PBH}^\mathrm{max}$. The PBH mass $M_\mathrm{PBH}=\gamma\left(\frac43\pi H_\mathrm{PBH}^{-3}\right)(3M_\mathrm{Pl}^2H_\mathrm{PBH}^2)$ collapsed from the Hubble-scale horizon mass at $\bar{t}_\mathrm{PBH}$ reads
\begin{align}\label{eq:PBHMass}
\frac{M_\mathrm{PBH}}{M_\odot}&=4\pi\gamma\frac{M_\mathrm{Pl}}{M_\odot}\frac{M_\mathrm{Pl}/\Gamma_0^{1/4}}{H_\mathrm{PBH}/\Gamma_0^{1/4}},
\end{align}
where $\gamma_\mathrm{PBH}=0.2$~\cite{Carr:1975qj}, $M_\mathrm{Pl}/M_\odot=2.182\times10^{-39}$, and $\overline{H}_\mathrm{PBH}\equiv H_\mathrm{PBH}/\Gamma_0^{1/4}=\sqrt{\bar{\rho}_\mathrm{tot}(\bar{t}_\mathrm{PBH};\bar{t}_n)}$, while the remaining factor can be approximately estimated as
\begin{align}\label{eq:Gamma0}
\frac{\Gamma_0^{1/4}}{M_\mathrm{Pl}}=\left(\frac{\pi^2}{90}g_\mathrm{dof}\right)^{1/2}\left(\frac{T_*}{M_\mathrm{Pl}}\right)^2\frac{\bar{t}_*}{\bar{t}_\mathrm{nuc}}e^{-W\left(\frac{\bar{\beta}}{8}\right)}
\end{align}
by recalling $\Gamma_0^{1/4}\equiv H_\mathrm{nuc}e^{-\beta/(8H_\mathrm{nuc})}$ with $3M_\mathrm{Pl}^2H_\mathrm{nuc}^2\approx(\pi^2/30)g_\mathrm{dof}T_\mathrm{nuc}^4$, $(T_\mathrm{nuc}/T_*)^2=\bar{t}_*/\bar{t}_\mathrm{nuc}$, and $\beta/(8H_\mathrm{nuc})\approx W(\bar{\beta}/8)$. Here $\bar{t}_*$ is solved from $F(\bar{t}_*;\bar{t}_n)=0.7$ and $\bar{t}_\mathrm{nuc}\approx(4/\bar{\beta})W(\bar{\beta}/8)$ is determined from $\bar{\beta}$ that can be further determined from given $\alpha_*$ and $(\beta/H)_*$. Note here that $T_*$ is another input parameter in addition to the strength factor $\alpha_*$ and inverse duration $(\beta/H)_*$ for estimating the PBH mass function. For the most probable case of PBH formations with $\bar{t}_n=\bar{t}_n^\mathrm{min}$ and $\bar{t}_\mathrm{PBH}=\bar{t}_\mathrm{PBH}^\mathrm{max}$, we have found a ready-to-use fitting formula for the PBH mass,
\begin{align}\label{eq:FitPBHM}
M_\mathrm{PBH}=3M_\odot&\left(\frac{g_\mathrm{dof}(0.1\,\mathrm{GeV})}{g_\mathrm{dof}(T_*)}\right)^\frac12\left(\frac{T_*}{0.1\,\mathrm{GeV}}\right)^{-2}\nonumber\\
&\times\mathrm{Fit}_M\left(\frac{\alpha_*}{0.6},\frac{(\beta/H)_*}{3}-1\right),
\end{align}
where the last factor in the fitting formula admits a simple form from combinations of power-law and exponential terms as
\begin{align}
\mathrm{Fit}_M(A,B)=\frac13+\frac16e^{-\frac{B}{2}}+\frac{A^{-0.55}}{2}-\frac{A^{-0.55}}{100}e^{-\frac{B}{2}}.
\end{align}

\begin{figure*}
    \centering
    \includegraphics[width=0.8\textwidth]{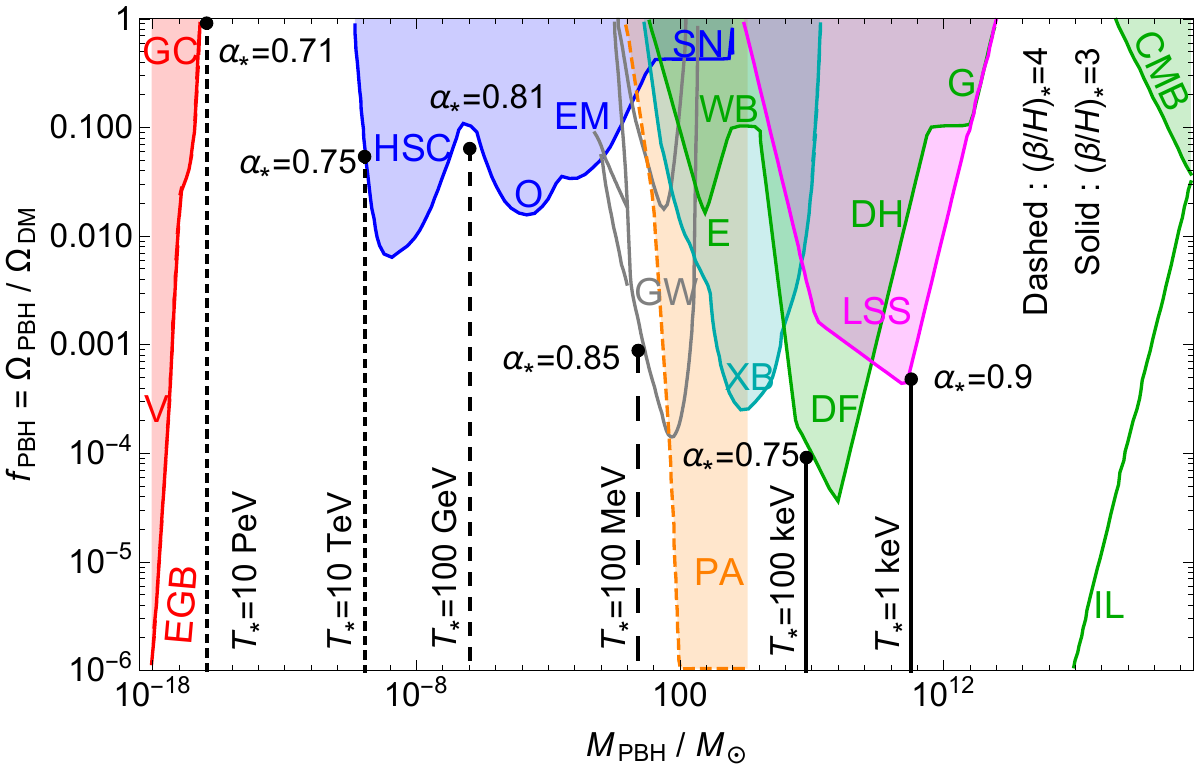}
    \caption{Illustrative examples of FOPTs at different percolation temperatures $T_*$ with different strength factors $\alpha_*$ and inverse durations $(\beta/H)_*=4$ (vertical dashed lines) and $(\beta/H)_*=3$ (vertical solid lines), in which cases the associate PBH masses and abundances roughly reach the boundaries of current PBH constraints from the collection~\cite{Carr:2020xqk} (and references therein but with GW constraints replaced by recent updates~\cite{Hutsi:2020sol,Nitz:2022ltl,Chen:2021nxo}). Note that the micro-lensing constraints from Subaru/HSC Andromeda observations~\cite{Niikura:2017zjd} and Optical Gravitational Lensing Experiment (OGLE)~\cite{Niikura:2019kqi} (recently updated in Refs.~\cite{Mroz:2024wag,Mroz:2024mse} for OGLE III+OGLE IV) could be significantly enhanced for PBHs dressed with dark matter minihalos~\cite{Cai:2022kbp}.}
    \label{fig:PBHExamples}
\end{figure*}

The PBH abundance $f_\mathrm{PBH}=(a_\mathrm{eq}/a_\mathrm{PBH})\Omega_\mathrm{PBH}/\Omega_\mathrm{DM}^\mathrm{eq}$ normalized to the dark matter fraction $\Omega_\mathrm{DM}^\mathrm{eq}=0.42$ at the matter-radiation equality can be estimated from noting that the PBH fraction $\Omega_\mathrm{PBH}(\bar{t}_\mathrm{PBH})$ at formation time is exactly the probability $P(t_n)$ for a Hubble volume $V_H(t)=\frac43\pi H(t)^{-3}$ not to decay until the delayed-decayed nucleation time $\bar{t}_n$. This can be further broken into infinite time intervals, $P(t_n)=\Pi_{t=t_i}^{t_n}\mathrm{d}P(t)$, each of which can be approximated as $\mathrm{d}P(t)=1-\Gamma(t)V_H(t)\mathrm{d}t\approx\exp[-\Gamma(t)V_H(t)\mathrm{d}t]$ for this Hubble volume not to decay within the time interval $\mathrm{d}t$ at the time $t$. Therefore, the PBH fraction at formation time reads
\begin{align}
\Omega_\mathrm{PBH}&(\bar{t}_\mathrm{PBH})=P(\bar{t}_n)=\exp\left[-\int_{t_i}^{t_n}\mathrm{d}t\Gamma(t)V_H(t)\right]\nonumber\\
&=\exp\left[-\frac43\pi\int_{\bar{t}_i}^{\bar{t}_n}\mathrm{d}\bar{t}\,e^{\bar{\beta}\bar{t}}\left(\frac{\sqrt{\bar{t}/\bar{t}_\mathrm{PBH}}}{\bar{H}_\mathrm{PBH}}\right)^3\right].
\end{align}
The remaining factor $a_\mathrm{eq}/a_\mathrm{PBH}=T_\mathrm{PBH}/T_\mathrm{eq}$ with $T_\mathrm{eq}\approx0.75$ eV can be estimated by approximating $3M_\mathrm{Pl}^2H_\mathrm{PBH}^2=(\pi^2/30)g_\mathrm{dof}T_\mathrm{PBH}^4$ after replacing $H_\mathrm{PBH}=\overline{H}_\mathrm{PBH}\Gamma_0^{1/4}$ with previously computed $\overline{H}_\mathrm{PBH}$ and $\Gamma_0^{1/4}/M_\mathrm{Pl}$, that is,
\begin{align}
\frac{T_\mathrm{PBH}}{M_\mathrm{Pl}}=\bar{\rho}_\mathrm{tot}^{1/4}(\bar{t}_\mathrm{PBH};\bar{t}_n)\left(\frac{T_*}{M_\mathrm{Pl}}\right)\left(\frac{\bar{t}_*}{\bar{t}_\mathrm{nuc}}\right)^\frac12e^{-\frac12W\left(\frac{\bar{\beta}}{8}\right)}.
\end{align}
It is worth noting that the PBH abundance admits no dependence on the relativistic effective degrees of freedom $g_\mathrm{dof}$ as it is canceled out in evaluating $T_\mathrm{PBH}/M_\mathrm{Pl}$. For the most probable case of PBH formations with $\bar{t}_n=\bar{t}_n^\mathrm{min}$ and $\bar{t}_\mathrm{PBH}=\bar{t}_\mathrm{PBH}^\mathrm{max}$, we have found a ready-to-use fitting formula for the PBH abundance,
\begin{align}\label{eq:FitPBHf}
f_\mathrm{PBH}=\frac{T_*}{T_\mathrm{eq}\Omega_\mathrm{DM}^\mathrm{eq}}\exp\left\{-\exp\left[\mathrm{Fit}_f\left(\frac{\alpha_*}{0.6},\frac{(\beta/H)_*}{3}-1\right)\right]\right\},
\end{align}
where the last factor in the fitting formula admits a simple form from combinations of power-law and linear terms as $\mathrm{Fit}_f(A,B)=3\left[\left(U+XA^{-m}\right)+\left(V+YA^{-n}\right)B\right]$ with $U=-0.93$, $V=0.32$, $X=2.11$, $Y=0.17$, $m=0.32$, and $n=0.81$. The goodness of fit for the PBH mass and abundance can be seen in Fig.~\ref{fig:FitPBH} with the fitting formulas shown in white dashed contours. 

A typical benchmark point reads $M_\mathrm{PBH}=3\,M_\odot$ and $f_\mathrm{PBH}=6\times10^{-7}$ for $\alpha_*=0.6$, $(\beta/H)_*=3$, and $T_*=0.1\,\mathrm{GeV}$, providing a promising PBH origin for the recent mass-gap observations from PSR J0514-4002E~\cite{Barr:2024wwl,Chen:2024joj} and GW230529 181500~\cite{LIGOScientific:2024elc} events, the later of which can meet the newly estimated abundance $f_\mathrm{PBH}\sim10^{-3}$~\cite{Huang:2024wse} if we adopt $\alpha_*=0.7$, $(\beta/H)_*=3$, and  $T_*=0.1\,\mathrm{GeV}$.  Since the PBH mass is almost sensitive to the FOPT scale alone, other typical PBH masses can be exemplified with different percolation temperatures $T_*$ as shown in Fig.~\ref{fig:PBHExamples}, where associated PBH abundances can touch the boundaries of current PBH constraints~\cite{Carr:2020xqk} (with GW constraints replaced by recent updates~\cite{Hutsi:2020sol,Nitz:2022ltl,Chen:2021nxo}) by tuning the strength factors $\alpha_*$ with respect to the inverse durations $(\beta/H)_*=4$ (vertical dashed lines) and $(\beta/H)_*=3$ (vertical solid lines) as the PBH abundance is more sensitive to $(\beta/H)_*$ than $\alpha_*$. Therefore, all PBH mass ranges including the asteroid-mass PBH as all dark matter,  the planet-mass PBH as weak lensing events, solar-mass PBH as LIGO-Virgo events, and supermassive black holes as in the galactic center can be easily realized with FOPTs at PeV, TeV, MeV, and keV scales.

\section{Curvature perturbations}\label{sec:PR}

The highly inhomogeneous evolution among radiations and vacuum energy densities would certainly induce curvature perturbations from the accumulated overdensity perturbations in different regions with different local nucleation times $\bar{t}_n$ of the first bubbles compared to the background with the earliest possible global nucleation time $\bar{t}_i$. In this section, we will estimate the curvature perturbations in real space by summing over all possible nucleation histories. For later convenience, we introduce a dimensionless version $\overline{V}_R(t)\equiv\Gamma_0^{3/4}V_R(t)$ of the physical volume at a time $t$ for a delayed-decayed region $V_R(t)=\frac43\pi(a(t)R)^3$ of a comoving size $R$, 
\begin{align}
\frac43\pi\left[\sqrt{\Gamma_0^{1/4}t}\left(C_0\Gamma_0^{1/8}R\right)\right]^3\equiv\frac43\pi\left(\sqrt{\bar{t}}\bar{R}\right)^3\equiv\overline{V}_{\bar{R}}(\bar{t}),
\end{align}
where the scale factor is dominated by radiation as $a(t)=C_0t^{1/2}$ recalling our effective approximation by transferring the inhomogeneities in scale factors temporarily into the equation of state, while the introduced dimensionless comoving scale $\bar{R}\equiv\left(C_0\Gamma_0^{1/8}R\right)$ serves as a convenient notation when the scale enters the Hubble horizon at $t_R$  by $\bar{R}_H^{-1}=a(t_R)H(t_R)/(C_0\Gamma_0^{1/8})=\sqrt{\bar{t}}_R\overline{H}(\bar{t}_R)=\sqrt{\bar{t}_R\bar{\rho}_\mathrm{tot}(\bar{t}_R;\bar{t}_i)}$. Then, we calculate the probability for this region $\overline{V}_{\bar{R}}(\bar{t})$ not to decay until a time $t'$, that is,
\begin{align}
P_R(t')&=\exp\left[-\int_{\bar{t}_i}^{\bar{t}'}\mathrm{d}\bar{t}\,e^{\bar{\beta}\bar{t}}\overline{V}_{\bar{R}}(\bar{t})\right]\\
&=\exp\left[\frac43\pi\bar{R}^3\bar{t}^{\frac52}E_{-\frac32}(-\bar{\beta}\bar{t})\bigg|_{\bar{t}_i}^{\bar{t}'}\right]\equiv\overline{P}_{\bar{R}}(\bar{t}'),
\end{align}
where $E_n(z)\equiv\int_1^{\infty}e^{-zt}t^{-n}\mathrm{d}t$ is the exponential integral.

Next, we define the smoothed density contrast field at a comoving scale $R$ by $\delta_R(t,\mathbf{x})=\int\mathrm{d}^3\mathbf{x}'W_R(|\mathbf{x}-\mathbf{x}'|)\delta(t,\mathbf{x})$ with some window function $W_R(r)=3\Theta(R-r)/(4\pi R^3)$. To calculate the variance of density perturbations at a comoving smoothing scale $R$, $\sigma_\delta^2(t;R)\equiv\langle(\delta_R(t,\mathbf{x})-\langle\delta_R(t,\mathbf{x})\rangle)^2\rangle=\langle\delta_R^2(t,\mathbf{x})\rangle-\langle\delta_R(t,\mathbf{x})\rangle^2$, recall that the spatial dependence in the density $\rho(t,\mathbf{x})=\rho(t;t_i+\Delta t(\mathbf{x}))\equiv\rho(t;t_n)$ comes from different local nucleation times of the first bubble in that region compared to the background with the earliest possible global nucleation time $\bar{t}_i$, and hence the average value for the overdensity $\delta(t,\mathbf{x})\equiv\rho(t;t_n)/\rho(t;t_i)-1\equiv\delta(t;t_n)$ at some comoving smoothing scale $R$ can be calculated by summing over all regions of size $R$ that had not yet decayed until the time considered,
\begin{align}
\langle\delta_{R}(t,\mathbf{x})\rangle_{\mathbf{x}}&\equiv\langle\delta_R(t; t_i+\Delta t(\mathbf{x}))\rangle_{\mathbf{x}\in\mathbb{R}^3/R^3}\equiv\langle\delta(t;t_n)\rangle_{t_n}\nonumber\\
&=\int_{t_i}^{U(t)}\delta(t;t')P_R(t')\Gamma(t')V_R(t')\mathrm{d}t',
\end{align}
where $\mathbb{R}^3/R^3$ is a quotient set differed by a scale $R$, while the upper bound for the nucleation time integration reads
\begin{align}
U(t)=\begin{cases}
t_n^\mathrm{max}, &\quad t<t_n^\mathrm{max}\equiv t_\mathrm{PBH}^\mathrm{min},\\
t_n(t), &\quad t_n^\mathrm{max}\equiv t_\mathrm{PBH}^\mathrm{min}<t<t_\mathrm{PBH}^\mathrm{max},\\
t_n^\mathrm{min}, &\quad t>t_\mathrm{PBH}^\mathrm{max}.
\end{cases}
\end{align}

The choice for the upper-bound function $U(t)$ is the key part of our proposal for the nucleation history summation, which merits further elaborations as below: (i) For the time considered smaller than the earliest possible time for PBH formations, $t<t_n^\mathrm{max}\equiv t_\mathrm{PBH}^\mathrm{min}$, all regions of size $R$ will contribute to the above integration, and hence the upper bound $U(t)=t_n^\mathrm{max}$ should cover all the possible delayed-decayed nucleation times regardless whether or not these regions can eventually collapse into PBHs. (ii) For the time considered within the possible time interval for PBH formations, $t_\mathrm{PBH}^\mathrm{min}<t<t_\mathrm{PBH}^\mathrm{max}$, all regions of size $R$ except for those regions that have already collapsed into PBHs should contribute to the above integration. Since regions that decay later would collapse into PBHs earlier, and hence we can solve the condition $\delta(t; t_n(t), t_i)=\delta_c$ for the upper bound $U(t)=t_n(t)$, beyond which those regions have already collapsed into PBHs by the time $t$. (iii) For the time considered larger than the latest possible time for PBH formation, all regions that can collapse into PBHs should be left out of the above integration, and hence the upper bound $U(t)=t_n^\mathrm{min}$ should only cover those regions with the local nucleation times not late enough to produce PBHs. The above procedure of nucleation history summation can also be used to calculate the average value for any function of the filtered density field as
\begin{align}
\langle f(\delta_{\bar{R}}(\bar{t}))\rangle=\int_{\bar{t}_i}^{U(\bar{t})}f(\delta(\bar{t};\bar{t}'))\overline{P}_{\bar{R}}(\bar{t}')e^{\bar{\beta}\bar{t}'}\overline{V}_{\bar{R}}(\bar{t}')\mathrm{d}\bar{t}'.
\end{align}

\begin{figure*}
    \centering
    \includegraphics[width=0.48\textwidth]{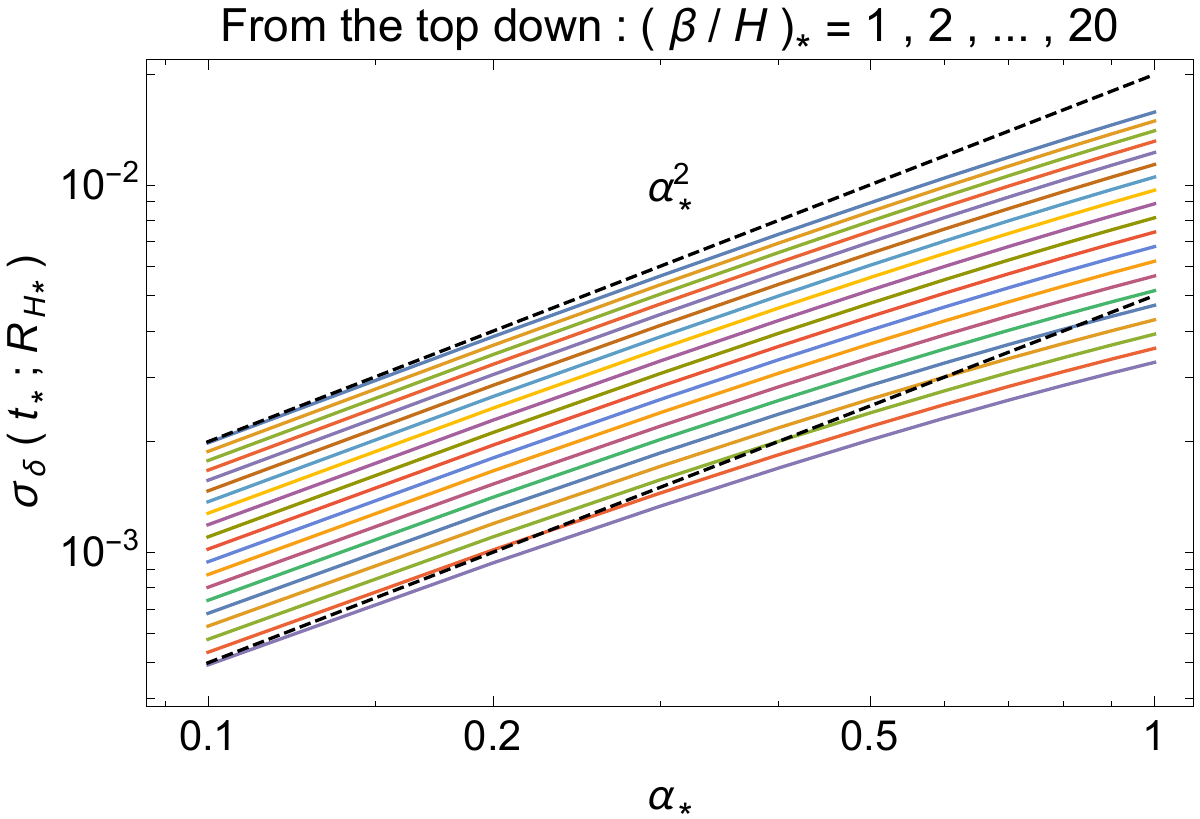}
    \includegraphics[width=0.48\textwidth]{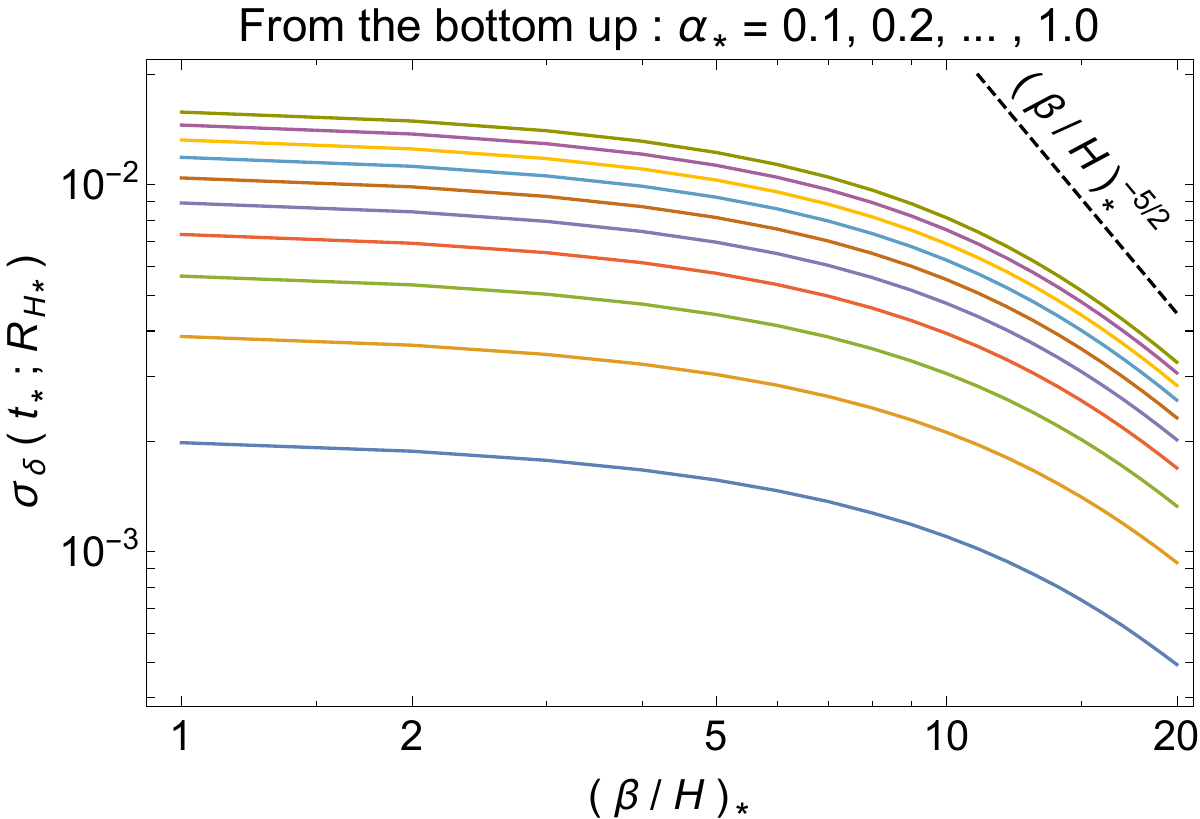}\\
    \includegraphics[width=0.48\textwidth]{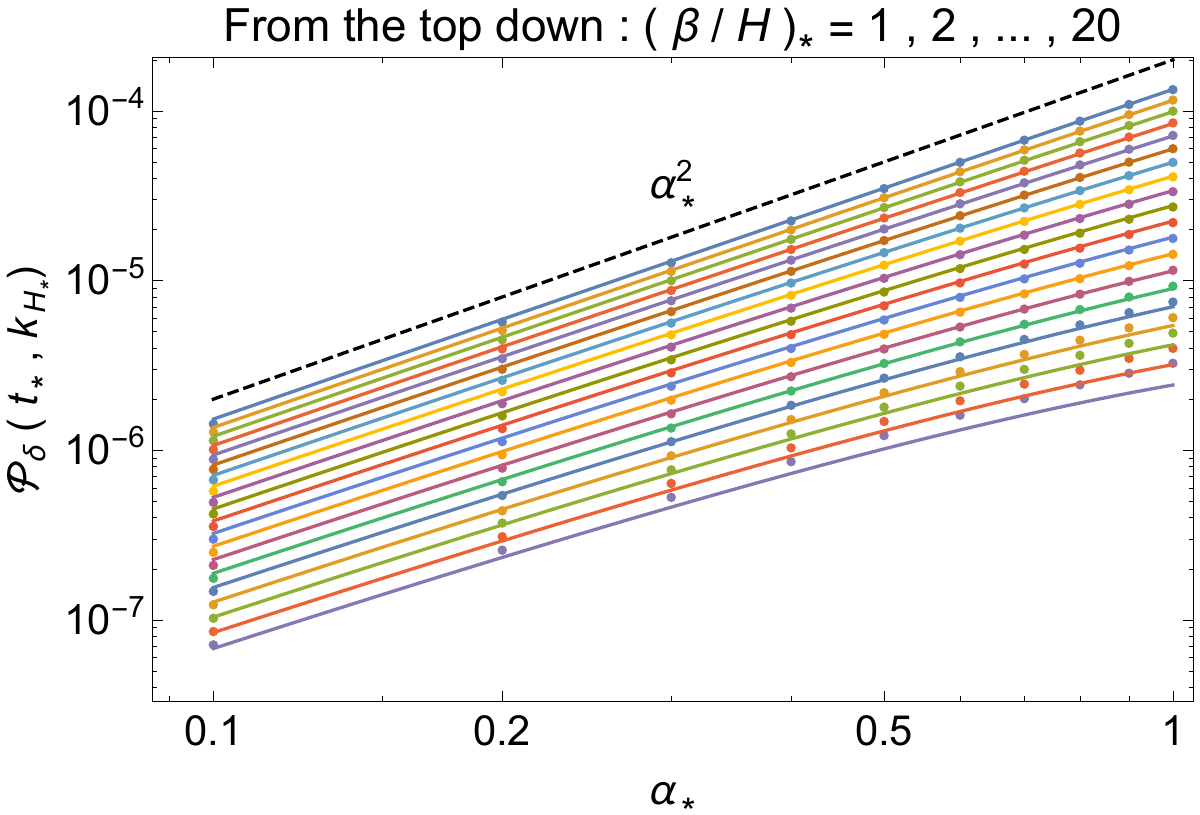}
    \includegraphics[width=0.48\textwidth]{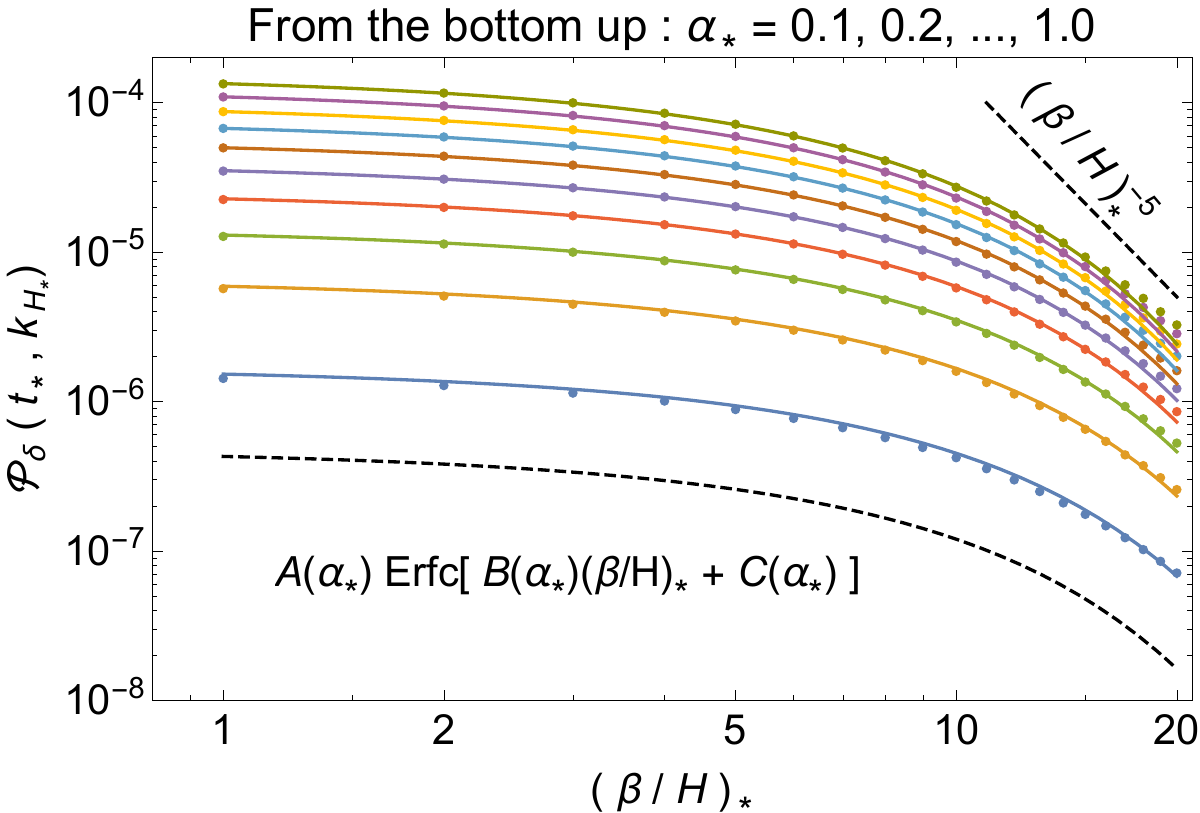}\\
    \caption{The standard deviation (top) and dimensionless power spectrum (bottom) for the accumulated overdensity perturbations filtered at a scale that enters the Hubble horizon around the bubble percolation time. The parameter dependences are shown separately with respect to the strength factor $\alpha_*$ (left) and inverse duration $(\beta/H)_*$ (right) for some fixed values of $(\beta/H)_*$ and $\alpha_*$, respectively.}
    \label{fig:SigmaDeltaStar}
\end{figure*}

Then, we can directly compute the variance of the accumulated overdensity field smoothed at a comoving scale $R$ by $\sigma_\delta^2(\bar{t};\bar{R})=\langle\delta_{\bar{R}}(\bar{t})^2\rangle-\langle\delta_{\bar{R}}(\bar{t})\rangle^2$, which can be particularly evaluated at a specific comoving scale $R_H^{-1}=a(t_R)H(t_R)$ that enters the Hubble horizon roughly around the percolation time $t_*$, that is $t_{R_H}=t_*$. The resulting standard deviation of the overdensity perturbation $\sigma_\delta(\bar{t}_*;\bar{R}_{H_*}=1/\sqrt{\bar{t}_*\bar{\rho}_\mathrm{tot}(\bar{t}_*;\bar{t}_i)})$ around bubble percolation time can be shown in the top row of Fig.~\ref{fig:SigmaDeltaStar}, whose parameter dependence on the strength factor $\alpha_*$ is almost linear and the parameter dependence on the inverse duration $(\beta/H)_*$ should asymptotic to power of $-5/2$ at the large inverse duration limit as argued in Refs~\cite{Liu:2022lvz,Elor:2023xbz}. This combined leads to the heuristic scaling $\sigma_\delta(\bar{t}_*;\bar{R}_{H_*})\sim\alpha_*(\beta/H)_*^{-5/2}$.

\begin{figure*}
    \centering
    \includegraphics[width=0.9\textwidth]{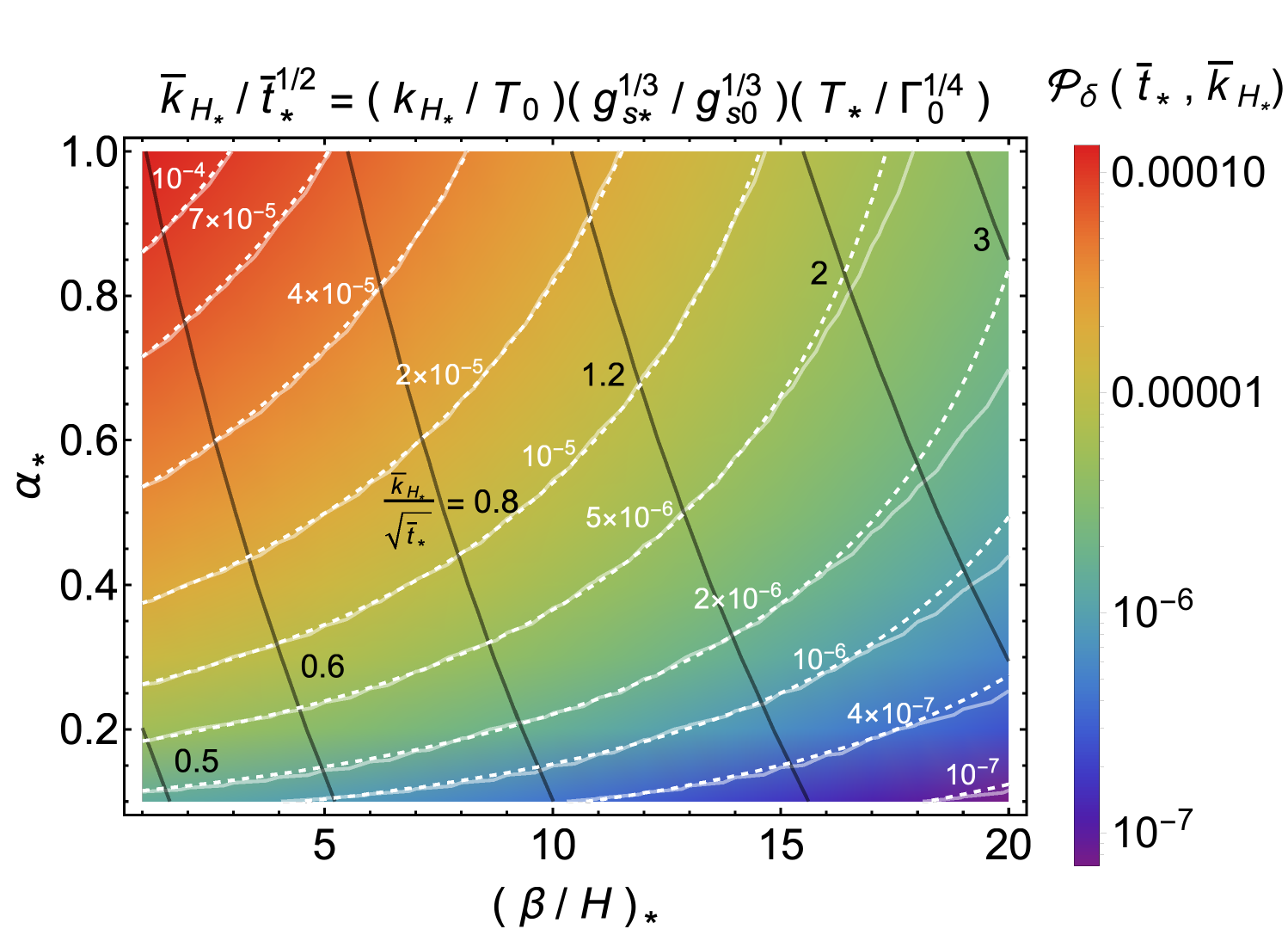}\\
     \includegraphics[width=0.49\textwidth]{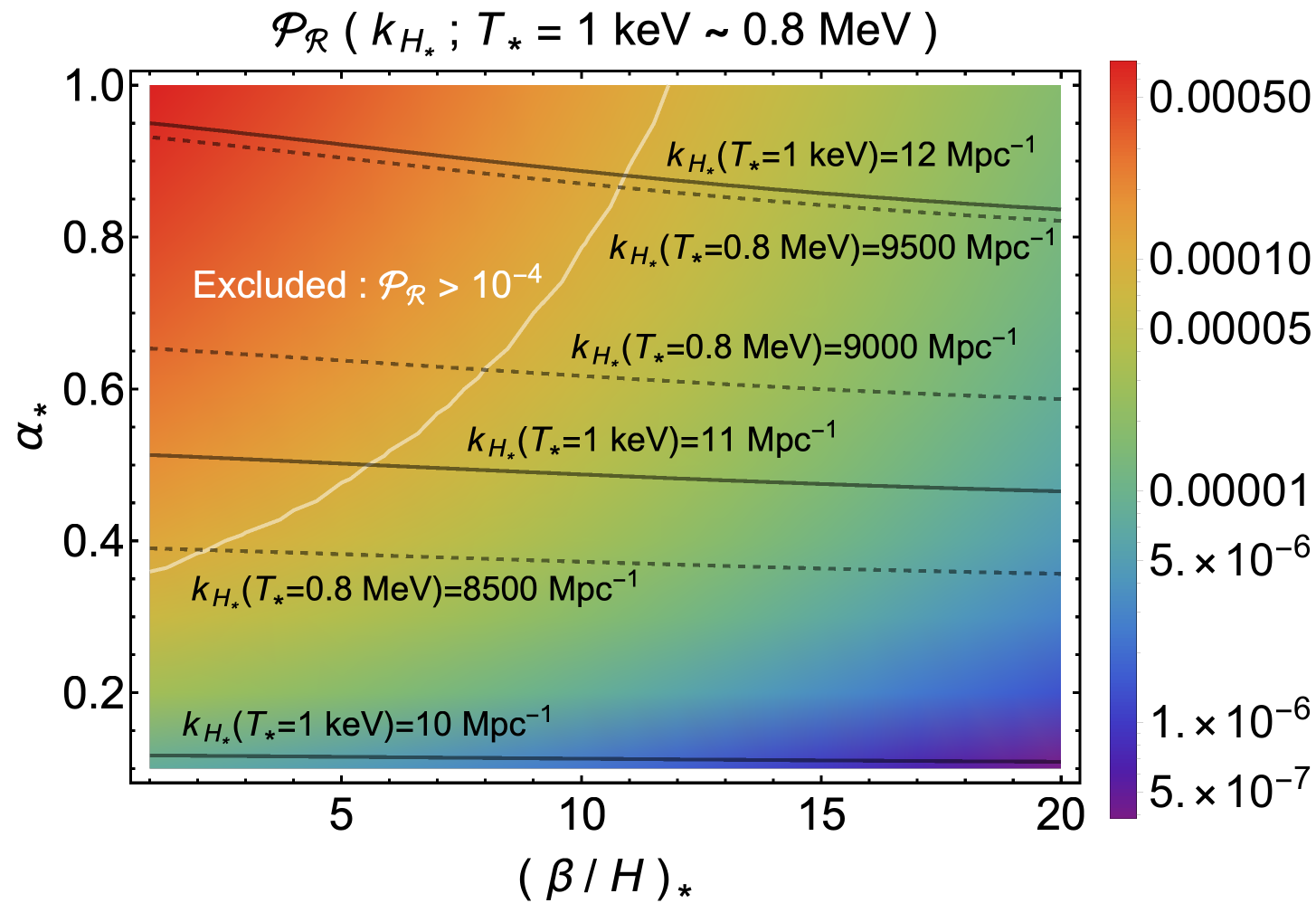}
      \includegraphics[width=0.49\textwidth]{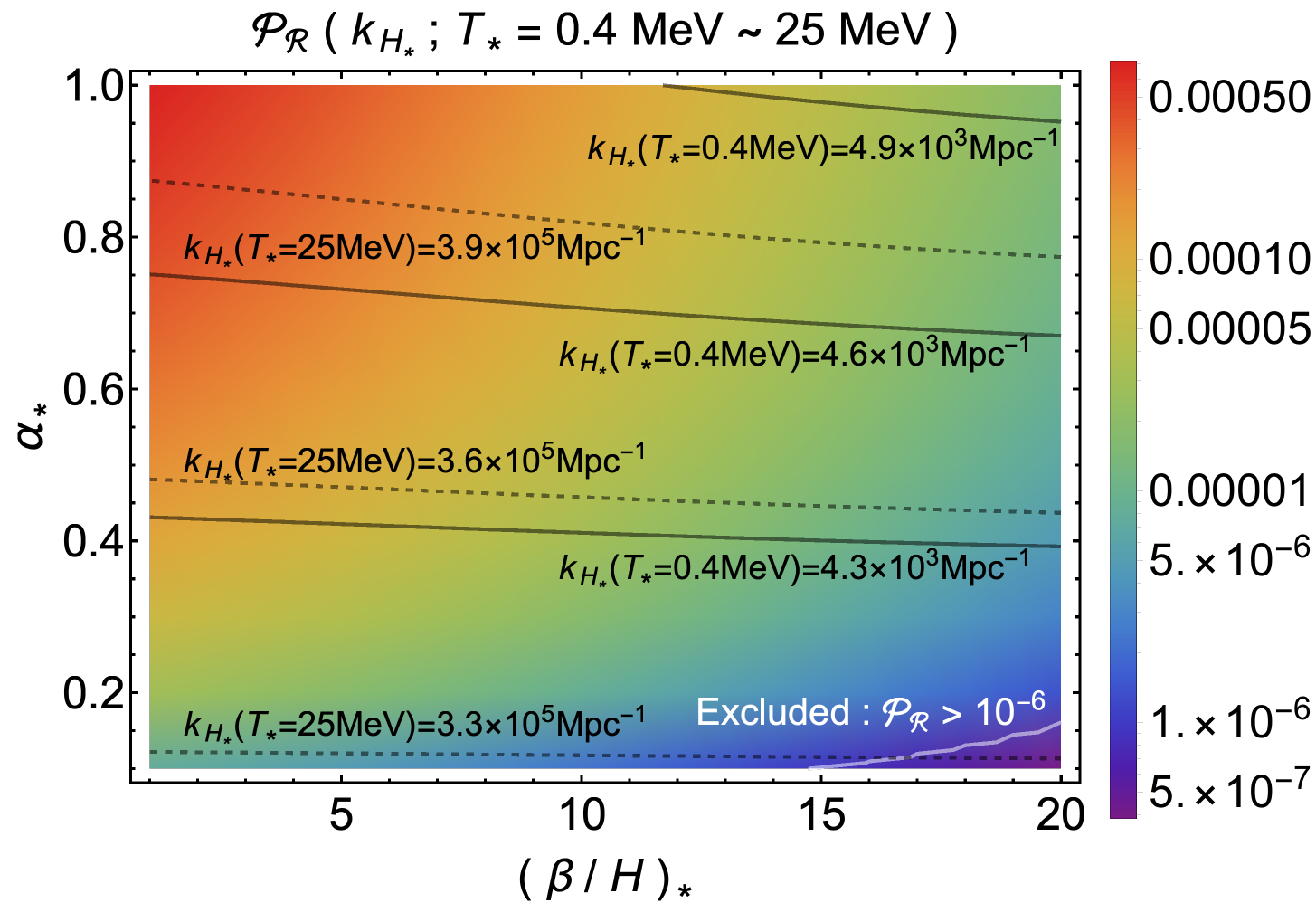}\\
    \caption{The power spectra for the accumulated overdensities (top panel) with the fitting formula shown in white dashed contours and the induced curvature perturbations (bottom panels) with corresponding ranges of percolation temperatures to manifest the observational exclusion constraints (white contours).}
    \label{fig:PowerSpectrum}
\end{figure*}

Finally, we can turn to calculate the power spectrum for the overdensity perturbations and the associated curvature perturbations. The basic strategy is to transform the definite integral of $\sigma_\delta^2(t;R)$ into a definite integral with a variable upper limit by a $k$-space top-hat window function,
\begin{align}
\sigma_\delta^2(t;R)=\int_0^\infty\frac{\mathrm{d}k}{k}\widetilde{W}_R^2(k)\mathcal{P}_\delta(t,k)
\approx\int_0^{k_\Lambda}\frac{\mathrm{d}k}{k}\mathcal{P}_\delta(t,k),
\end{align}
where the choice for the cut-off scale $k_\Lambda$ is not unique. A natural but naive choice would be $k_\Lambda=1/R$~\cite{Kimura:2021sqz,Wang:2022nml}. Another plausible choice $k_\Lambda=3\pi/(5R)$ can be obtained from requiring the same normalization $k_\Lambda\equiv\int_0^{k_\Lambda}\mathrm{d}k=\int_0^\infty\mathrm{d}k\,\widetilde{W}_R^2(k)=3\pi/(5R)$ for the Fourier transformed window function $\widetilde{W}_R(k)=3j_1(kR)/(kR)$.  The previous choice~\cite{Liu:2022lvz} for the window function with a Gaussian form $\widetilde{W}_R(k)=\exp(-k^2R^2/2)$ renders a much smaller cutoff $k_\Lambda\equiv\int_0^{k_\Lambda}\mathrm{d}k=\int_0^\infty\mathrm{d}k\,\widetilde{W}_R^2(k)=\sqrt{\pi}/(2R)$. We further test and present in this paper a relatively compromise choice $k_\Lambda=3\pi/(4R)$ to see the effect on density perturbations. The dimensionless power spectrum can be obtained by directly taking a scale derivative~\cite{Kimura:2021sqz,Wang:2022nml},
\begin{align}
\mathcal{P}_\delta(t,k)=\frac{\partial\sigma_\delta^2(t,k')}{\partial\ln k'}\bigg|_{k'=\frac{4k}{3\pi}},
\end{align}
where an abbreviation $\sigma_\delta^2(t;R\equiv 1/k)\equiv\sigma_\delta^2(t,k)$ as well as $\delta_{R\equiv1/k}(t)\equiv\delta(t,k)$ is introduced for later clarity. This scale derivative on the density variance can be further transferred into scale derivatives of average values by
\begin{align}
\frac{\partial\sigma_\delta^2(t,k)}{\partial\ln k}=\frac{\partial\langle\delta(t,k)^2\rangle}{\partial\ln k}-2\langle\delta(t,k)\rangle\frac{\partial\langle\delta(t,k)\rangle}{\partial\ln k},
\end{align}
where the scale derivative for the average value of an arbitrary function of the density contrast reads
\begin{align}
\frac{\partial\langle f(\delta(t,k))\rangle}{\partial\ln k}=\int_{\bar{t}_i}^{U(\bar{t})}f(\delta(\bar{t};\bar{t}'))e^{\bar{\beta}\bar{t}'}\frac{\partial}{\partial\ln k}\left[\overline{PV}(\bar{t}',\bar{k})\right]\mathrm{d}\bar{t}'.
\end{align}
Here we introduce $\overline{PV}(\bar{t},\bar{k})\equiv\overline{P}(\bar{t},\bar{k})\overline{V}(\bar{t},\bar{k})$ for short with abbreviations $\overline{P}_{\bar{R}\equiv1/\bar{k}}(\bar{t})\equiv\overline{P}(\bar{t},\bar{k})$ and $\overline{V}_{\bar{R}\equiv1/\bar{k}}(t)\equiv\overline{V}(\bar{t},\bar{R})$. Therefore, we can numerically obtain the dimensionless power spectrum $\mathcal{P}_\delta(\bar{t},\bar{k})$ for the accumulated overdensities from those delayed-decayed regions of any size and at any time, which can be specifically evaluated at a comoving scale that enters the Hubble horizon around the bubble percolation time, $\mathcal{P}_\delta(\bar{t}_*,\bar{k}_{H_*})$, as shown in dots in the bottom row of Fig.~\ref{fig:SigmaDeltaStar} with respect to the strength factor $\alpha_*$ and inverse duration $(\beta/H)_*$ for some fixed values of $(\beta/H)_*$ and $\alpha_*$ in the left and right panels, respectively. For a small $(\beta/H)_*$, $\mathcal{P}_\delta(\bar{t}_*,\bar{k}_{H_*})$ is almost quadratic in $\alpha_*$ but gradually deviates from that for a larger $(\beta/H)_*$. For a small $\alpha_*$, $\mathcal{P}_\delta(\bar{t}_*,\bar{k}_{H_*})$ aligns closely with a complementary error function in $(\beta/H)_*$ but also slowly deviates from that for a larger $\alpha_*$ at the large $(\beta/H)_*$ limit, which should be asymptotic to $(\beta/H)_*^{-5}$ as theoretically anticipated. For the parameter space we explore in this study, we found a good fitting formula as
\begin{align}
\mathcal{P}_\delta(\bar{t}_*,\bar{k}_{H_*})=A(\alpha_*)\mathrm{Erfc}\left[B(\alpha_*)(\beta/H)_*+C(\alpha_*)\right]
\end{align}
with $A(\alpha_*)=(0.00369\alpha_*+0.0185)^2\alpha_*^2$, $B(\alpha_*)=0.00665\alpha_*+0.0571$, and $C(\alpha_*)=0.278\alpha_*^{0.615}+0.436$, whose solely dependence on $\alpha_*$ and $(\beta/H)_*$ can be shown separately as solid curves in the left and right panels of the bottom row of Fig.~\ref{fig:SigmaDeltaStar}, respectively, while the full dependence on both $\alpha_*$ and $(\beta/H)_*$ is presented in the top panel of Fig.~\ref{fig:PowerSpectrum} with the fitting formula shown in dashed curves.

At last, we are able to compute the power spectrum for the induced curvature perturbations from the previously obtained power spectrum of accumulated overdensities. Since the density perturbations $\delta(t,k)$ can be related to the induced curvature perturbations $\mathcal{R}(t,k)\equiv\mathcal{T}(t,k)\mathcal{R}(k)$ by~\cite{Josan:2009qn}
\begin{align}
\delta(t,k)=\frac{2(1+w)}{5+3w}\left(\frac{k}{a(t)H(t)}\right)^2\mathcal{T}(t,k)\mathcal{R}(k)
\end{align}
in the radiation-dominated era with the equation of state parameter $w=1/3$ where the transfer function admits $\mathcal{T}(t,k)=3(\sqrt{3}aH/k)j_1(k/(\sqrt{3}aH))$, the corresponding power spectra can be related as
\begin{align}
\mathcal{P}_\delta(t,k)=\frac{16}{3}\left(\frac{k}{aH}\right)^2j_1\left(\frac{k}{\sqrt{3}aH}\right)^2\mathcal{P}_\mathcal{R}(k),
\end{align}
whose value at horizon crossing simply reads
\begin{align}
\mathcal{P}_\mathcal{R}(k_H)=\frac{3\mathcal{P}_\delta(t_{R_H},k_H)}{16j_1(1/\sqrt{3})^2}.
\end{align}
A useful numeric trick for this special value of the spherical Bessel function $j_1(1/\sqrt{3})\approx3/16$ serves as a good approximation with better than $1\%$ accuracy. Therefore, the final dimensionless power spectrum of curvature perturbations induced by accumulated overdensities from delayed-decayed patches can be estimated as 
\begin{align}
\mathcal{P}_\mathcal{R}(k_{H_*})\approx\frac{16}{3}\mathcal{P}_\delta(t_*,k_{H_*})
\end{align}
at a comoving scale entering the Hubble horizon around the bubble percolations. However, the previously obtained numerical result for $\mathcal{P}_\delta(\bar{t}_*,\bar{k}_{H_*})$ shown in the top panel of Fig.~\ref{fig:PowerSpectrum} behaves as a function of the strength factor $\alpha_*$ and inverse duration $(\beta/H)_*$, while both dimensionless time $\bar{t}_*$ and the dimensionless scale $\bar{k}_{H_*}$ also depend on $\alpha_*$ and $(\beta/H)_*$, so does their ratio $\bar{k}_{H_*}/\sqrt{\bar{t}_*}$ as shown in black contours. In fact, this ratio can be directly related to the dimensional comoving scale $k_{H_*}$ via the percolation temperature $T_*$ as
\begin{align}
k_{H_*}\equiv\bar{k}_{H_*}(C_0\Gamma_0^{1/8})=\frac{\bar{k}_{H_*}}{\sqrt{\bar{t}_*}}\left(\frac{g_{s0}}{g_{s*}}\right)^\frac13\frac{\Gamma_0^{1/4}/M_\mathrm{Pl}}{T_*/M_\mathrm{Pl}}T_0,
\end{align}
where we have inserted $C_0=(T_0/T_*)(g_{s0}/g_{s*})^{1/3}t^{-1/2}$ from the entropy conservation $g_{s*}T_*^3a(t_*)^3=g_{s0}T_0^3a_0^3$ with $a(T_*)=C_0t_*^{1/2}$ in the radiation dominated era, and $T_0=2.73\,\mathrm{K}$ and $g_{s0}=3.94$ are the cosmic microwave background (CMB) temperature and the effective number of degrees of freedom in entropy at present day, respectively, while $\Gamma_0/M_\mathrm{Pl}$ factor can be evaluated with~\eqref{eq:Gamma0} given $\alpha_*$, $(\beta/H)_*$, and $T_*$. Therefore, for given percolation temperature $T_*$, the dimensionless comoving scale $k_{H_*}$ can be mapped onto the same $\alpha_*-(\beta/H)_*$ plane where $\mathcal{P}_\delta(\bar{t}_*,\bar{k}_{H_*})$ is calculated. Now, we can draw various observational exclusion curves for $\mathcal{P}_\mathcal{R}(k)$ within the range of scales of observations. For example, in the bottom panels of Fig.~\ref{fig:PowerSpectrum}, we have located two illustrative ranges of percolation temperatures so that the inferred comoving scale $k_{H_*}$ within the $\alpha_*-(\beta/H)_*$ parameter space considered could cover the corresponding observational ranges where the observational constraints on the upper bound of $\mathcal{P}_\mathcal{R}(k)$ are imposed, that is, the CMB spectral $y$-distortion and $\mu$-distortion constraints on $\mathcal{P}_\mathcal{R}(k)\lesssim10^{-4}$ in the range $1\,\mathrm{Mpc}^{-1}\lesssim k\lesssim10^4\,\mathrm{Mpc}^{-1}$~\cite{Chluba:2012we,Chluba:2015bqa,Lucca:2019rxf} and the  ultra-compact minihalo (UCMH) abundance constraint $\mathcal{P}_\mathcal{R}(k)\lesssim10^{-6}$ in the range $4\times10^3\,\mathrm{Mpc}^{-1}\lesssim k\lesssim4\times10^5\,\mathrm{Mpc}^{-1}$~\cite{Clark:2015sha,Clark:2015tha}. Note that the current constraints on low-scale FOPTs from the above two observational bounds are slightly weaker than what we have obtained in our previous study~\cite{Liu:2022lvz}, which could be traced back to different window functions. More carefully investigations are certainly needed in future for curvature-perturbation constraints on these low-scale FOPTs.

\section{Conclusions and discussions}\label{sec:condis}

The interplay between the cosmological FOPTs and PBHs is of great interest to recent studies, among which the renewed attention on producing PBHs and curvature perturbations from cosmological FOPTs has been in hot debate. In this paper, we have revisited and elaborated our previously proposed accumulating mechanism for the overdensities produced in the delayed-decayed false vacuum island regions during the asynchronous progress of cosmological FOPTs. In particular, model-independent fitting formulas are given for future model-building investigations. Nevertheless, the current treatments are still far from satisfaction, which can be further improved in future studies from the following considerations:

First, numerical simulations are still needed to eventually settle the debates on whether the delayed-decayed regions could really lead to local overdensities, and whether these local overdensities could eventually collapse into PBHs according to the usual critical collapse criteria, and if not, whether they could eventually source curvature perturbations. The difficulty in carrying out such numerical simulations lies in the ultra-low probability of locating such delayed-decayed patches, requiring extremely large (and hence expensive) simulation volume to actually emerge such regions.

Second, if the first issue can be resolved and confirmed, the next challenge comes from analytically modeling the inhomogeneous evolution in completing the FOPT. Not only the full equations of motions are coupled integro-differential equations if the global background evolution is presumed \textit{a priori}, but also the definition of overdensity relies on the averaged density over all possible nucleation histories that in turn requires the solved evolution of overdensities. In this study, we simply adopt a brute strategy to approximate the global background as in the radiation-dominated era and the averaged background as those patches with the earliest possible nucleation time since they greatly outnumber the others. Future investigations should be rigorously implemented either with iterative method~\cite{Kanemura:2024pae} or transforming the two-dimensional system of integro-differential equation into a seven-dimensional system of first-order ordinary differential equations~\cite{Flores:2024lng}.

Third, if the above two issues can both be resolved, then the PBH mass function and induced curvature perturbations are straightforward to calculate with our proposals. The final task is to find model-independent fitting formulas for different types of cosmological FOPTs with nucleation rates of constant and Dirac-delta forms as well as exponentially linear or quadratic in time. 

Last, although our mechanism for producing PBHs and density perturbations is independent of details of particle physics models, the precise calculations, however, do reply on the underlying phase transition models. For example, the bubble wall velocity in the comoving radius of true vacuum bubbles from computing the false vacuum fraction has been fixed to be the speed of light. Lower wall velocity would further delay the percolation time and hence allow for later delayed-decayed nucleation time of the first bubble in those Hubble patches that collapse into smaller PBHs at an earlier time with higher abundances as shown qualitatively in Ref.~\cite{He:2023ado}. Model dependency would further come into play if one further takes into account the full time evolution of the wall velocity $v_w(t)$ before reaching its terminal value~\cite{Cai:2020djd} due to complex interactions between the bubble wall and thermal plasma. Another example is the strength factor of the phase transition defined as $\alpha(t)\equiv\Delta V(t)/\rho_r(t; t_i)$ usually evaluated at the percolation time where the vacuum energy density difference  $\Delta V(T)\equiv V_+-V_-(T)$ has been simply fixed at a constant value. In fact, for the slow FOPT with $1\lesssim(\beta/H)_*\lesssim20$ considered in this paper, the true vacuum potential energy density $V_-(T)$ could decrease significantly with decreasing temperature~\cite{Cai:2017tmh}, enlarging the vacuum energy difference that accumulates over time in those delayed-decayed patches, whose overdensities would more rapidly saturate the PBH threshold with less delayed-decayed time, resulting in smaller PBH mass with larger abundance. We will leave these model dependencies in the time evolution of the bubble wall velocity $v_w(t)$ and true-vacuum energy density $V_-(T(t))$ in future work of specific particle physics models of cosmological FOPTs.

\begin{acknowledgments}
We thank the helpful discussions with Ligong Bian, Jing Liu, Shi Pi, Misao Sasaki, Ke-Pan Xie, and the insightful correspondence with Yann Gouttenoire.
This work is supported by the National Key Research and Development Program of China Grant No. 2021YFC2203004, No. 2021YFA0718304,  and No. 2020YFC2201502,
the National Natural Science Foundation of China Grants No. 12105344, No. 12235019, No. 11821505, No. 11991052, and No. 11947302,
the Strategic Priority Research Program of the Chinese Academy of Sciences (CAS) Grant No. XDB23030100, No. XDA15020701, 
the Key Research Program of the CAS Grant No. XDPB15, 
the Key Research Program of Frontier Sciences of CAS,
and the Science Research Grants from the China Manned Space Project with No. CMS-CSST-2021-B01.
\end{acknowledgments}


\bibliography{ref}

\end{document}